\documentclass{aa}
\usepackage[T1]{fontenc}
\usepackage{amstext}
\usepackage{amsmath}
\usepackage{graphicx}
\usepackage{academicons,xcolor}
\usepackage[varg]{txfonts}  
\bibpunct{(}{)}{;}{a}{}{,}  
\usepackage[colorlinks, citecolor=blue, linkcolor=blue]{hyperref}

\definecolor{orcidlogocol}{HTML}{A6CE39}

\makeatletter

\iftrue
    \usepackage{twoopt}
    \usepackage{xstring}
    \newcommand{\adshref}[2]{\StrRight{#1}{19}[\adsid]\href{http://adsabs.harvard.edu/abs/\adsid}{#2}}
    
    \let\orgcitep\citep
    \let\orgcitet\citet
    \let\orgcitealt\citealt
    \let\orgcitealp\citealp
    \let\orgciteauthor\citeauthor

    \renewcommandtwoopt{\cite}[3][][]{\adshref{#3}
        {\def\hyper@linkstart##1##2{}
        \let\hyper@linkend\@empty\orgcitet[#1][#2]{#3}}}
    \renewcommandtwoopt{\citep}[3][][]{\adshref{#3}
        {\def\hyper@linkstart##1##2{}
        \let\hyper@linkend\@empty\orgcitep[#1][#2]{#3}}}
    \renewcommandtwoopt{\citet}[3][][]{\adshref{#3}
        {\def\hyper@linkstart##1##2{}
        \let\hyper@linkend\@empty\orgcitet[#1][#2]{#3}}}
    \renewcommandtwoopt{\citealt}[3][][]{\adshref{#3}
        {\def\hyper@linkstart##1##2{}
        \let\hyper@linkend\@empty\orgcitealt[#1][#2]{#3}}}
    \renewcommandtwoopt{\citealp}[3][][]{\adshref{#3}
        {\def\hyper@linkstart##1##2{}
        \let\hyper@linkend\@empty\orgcitealp[#1][#2]{#3}}}
    \renewcommandtwoopt{\citeauthor}[3][][]{\adshref{#3}
        {\def\hyper@linkstart##1##2{}
        \let\hyper@linkend\@empty\orgciteauthor[#1][#2]{#3}}}
    \newcommandtwoopt{\citeyearads}[3][][]
        {\href{http://adsabs.harvard.edu/abs/#3}
        {\def\hyper@linkstart##1##2{}
        \let\hyper@linkend\@empty\citeyear[#1][#2]{#3}}}
   \iftrue
        \renewcommandtwoopt{\cite}[3][][]{\adshref{#3}{\orgcitet[#1][#2]{#3}}}
        \renewcommandtwoopt{\citep}[3][][]{\adshref{#3}{\orgcitep[#1][#2]{#3}}}
        \renewcommandtwoopt{\citet}[3][][]{\adshref{#3}{\orgcitet[#1][#2]{#3}}}
        \renewcommandtwoopt{\citealt}[3][][]{\adshref{#3}{\orgcitealt[#1][#2]{#3}}}
        \renewcommandtwoopt{\citealp}[3][][]{\adshref{#3}{\orgcitealp[#1][#2]{#3}}}
    \fi
\fi

\def\instrefs#1{{\def\scsep{\def\scsep{,}}\@for\w:=#1\do{\scsep\ref{inst:\w}}}}
\renewcommand{\inst}[1]{\unskip$^{\instrefs{#1}}$}

\usepackage{xspace}

\renewcommand*\aa@pageof{, page \thepage{} of \pageref*{LastPage}} 

\usepackage{scalerel}
\usepackage{tikz}
\usetikzlibrary{svg.path}
\definecolor{orcidlogocol}{HTML}{A6CE39}
\tikzset{
  orcidlogo/.pic={
    \fill[orcidlogocol] svg{M256,128c0,70.7-57.3,128-128,128C57.3,256,0,198.7,0,128C0,57.3,57.3,0,128,0C198.7,0,256,57.3,256,128z};
    \fill[white] svg{M86.3,186.2H70.9V79.1h15.4v48.4V186.2z}
                 svg{M108.9,79.1h41.6c39.6,0,57,28.3,57,53.6c0,27.5-21.5,53.6-56.8,53.6h-41.8V79.1z M124.3,172.4h24.5c34.9,0,42.9-26.5,42.9-39.7c0-21.5-13.7-39.7-43.7-39.7h-23.7V172.4z}
                 svg{M88.7,56.8c0,5.5-4.5,10.1-10.1,10.1c-5.6,0-10.1-4.6-10.1-10.1c0-5.6,4.5-10.1,10.1-10.1C84.2,46.7,88.7,51.3,88.7,56.8z};
  }
}
\newcommand\orcidicon[1]{\href{https://orcid.org/#1}{\mbox{\scalerel*{
\begin{tikzpicture}[yscale=-1,transform shape]
\pic{orcidlogo};
\end{tikzpicture}
}{|}}}}

\onecolumn

\begin{document}

\lefthyphenmin=3

\title{Teegarden's Star revisited}
\subtitle{A nearby planetary system with at least three planets\thanks{Based on observations collected at the European Southern Observatory under ESO programme(s) 0103.C-0152(A). Table A.1 is only available in electronic form at the CDS via anonymous ftp to cdsarc.u-strasbg.fr (130.79.128.5) or via http://cdsweb.u-strasbg.fr/cgi-bin/qcat?J/A+A/.}}

\titlerunning{Teegarden's Star }
\authorrunning{S. Dreizler et al.}

\author{
    S.~Dreizler\inst{iag}
    \and R.~Luque\inst{uchicago,iaa}
    \and I.~Ribas\inst{ice,ieec}
    \and V.~Koseleva\inst{iag}
    \and H.\,L.~Ruh\inst{iag}
    \and E.~Nagel\inst{iag}
    \and F.\,J.~Pozuelos\inst{iaa,uliege,sst}
    \and M.~Zechmeister\inst{iag}
    \and A.~Reiners\inst{iag}
    \and J.\,A.~Caballero\inst{cab}
    \and P.\,J.~Amado\inst{iaa}
    \and V.\,J.\,S.~B\'ejar\inst{iac,ull}
    \and J.\,L.~Bean\inst{uchicago}
    \and M.~Brady\inst{uchicago}
    \and C.~Cifuentes\inst{cab}
    \and M.\,Gillon\inst{uliege}
    \and A.\,P.~Hatzes\inst{tls}
    \and Th.~Henning\inst{mpia}
    \and D.~Kasper\inst{uchicago}
    \and D.~Montes\inst{ucm}
    \and J.\,C.~Morales\inst{ice,ieec}
    \and  C.\,A.\,Murray\inst{colorado}
    \and E.\,Pallé\inst{iac,ull}
    \and A.~Quirrenbach\inst{lsw}
    \and A.~Seifahrt\inst{uchicago}
    \and A.~Schweitzer\inst{hs}
    \and J.~St\"urmer\inst{lsw}
    \and G.~Stef\'ansson\inst{princeton}
    \and J.\,I.~Vico~Linares\inst{caha}
}

\institute{
    \label{inst:iag}Institut f\"ur Astrophysik, Georg-August-Universit\"at, Friedrich-Hund-Platz 1, 37077 G\"ottingen, Germany\\
    \email{dreizler@astro.physik.uni-goettingen.de}
    \and \label{inst:uchicago} Department of Astronomy \& Astrophysics, University of Chicago, Chicago, IL 60637, USA
    \and \label{inst:iaa} Instituto de Astrofísica de Andalucía (IAA-CSIC), Glorieta de la Astronomía s/n, 18008 Granada, Spain
    \and \label{inst:ice}Institut de Ci\`encies de l'Espai (ICE, CSIC), Campus UAB, C/ Can Magrans s/n, 08193 Bellaterra, Spain
    \and \label{inst:ieec}Institut d'Estudis Espacials de Catalunya, C/~Gran Capit\`a, 2-4, 08028 Barcelona, Spain
    \and \label{inst:uliege}Astrobiology Research Unit, Universit\'e de Li\`ege, 19C All\'ee du 6 Ao\^ut, 4000 Li\`ege, Belgium
    \and \label{inst:sst}Space Sciences, Technologies and Astrophysics Research (STAR) Institute, Université de Liège, Allée du 6 Août 19C, 4000 Liège, Belgium 
    \and \label{inst:cab}Centro de Astrobiolog\'ia (CSIC-INTA), ESAC campus, Camino bajo del castillo s/n, 28692 Villanueva de la Ca\~nada, Madrid, Spain
    \and \label{inst:iac}Instituto de Astrof\'isica de Canarias, Av.~V\'ia L\'actea s/n, 38205 La Laguna, Tenerife, Spain
    \and \label{inst:ull}Departamento de Astrof\'isica, Universidad de La Laguna, 38206 La Laguna, Tenerife, Spain
    \and \label{inst:tls}Th\"uringer Landessternwarte Tautenburg, Sternwarte 5, 07778 Tautenburg, Germany 
    \and \label{inst:mpia}Max-Planck-Institut f\"ur Astronomie, K\"onigstuhl 17, 69117 Heidelberg, Germany 
    \and \label{inst:ucm}Departamento de F\'{i}sica de la Tierra y Astrof\'{i}sica and IPARCOS-UCM (Instituto de F\'{i}sica de Part\'{i}culas y del Cosmos de la UCM), Facultad de Ciencias F\'{i}sicas, Universidad Complutense de Madrid, 28040, Madrid, Spain 
    \and \label{inst:colorado}Department of Astrophysical and Planetary Sciences, University of Colorado Boulder, Boulder, CO 80309, USA
    \and \label{inst:lsw}Landessternwarte, Zentrum f\"ur Astronomie der Universit\"at Heidelberg, K\"onigstuhl 12, 69117 Heidelberg, Germany 
    \and \label{inst:hs}Hamburger Sternwarte, Universit\"at Hamburg, Gojenbergsweg 112, 21029 Hamburg, Germany
    \and \label{inst:princeton}Department of Astrophysical Sciences, Princeton University, 4 Ivy Lane, Princeton, NJ 08540, US
    \and \label{inst:caha}Centro Astron\'omico Hispano-Alem\'an, Observatorio de Calar Alto, Sierra de los Filabres, 04550 G\'ergal, Spain
}

\abstract{The two known planets in the planetary system of Teegarden's Star are among the most Earth-like exoplanets currently known. Revisiting this nearby planetary system with two planets in the habitable zone aims at a more complete census of planets around very low-mass stars. A significant number of new radial velocity measurements from CARMENES, ESPRESSO, MAROON-X, and HPF, as well as photometry from {\it TESS} motivated a deeper search for additional planets. We confirm and refine the orbital parameters of the two know planets Teegarden's Star b and c. We also report the detection of a third planet d with an orbital period of 26.13$\pm 0.04$\,d and a minimum mass of 0.82$\pm0.17$\,M$_\oplus$. A signal at 96\,d is attributed to the stellar rotation period. The interpretation of a signal at 172\,d remains open. The {\it TESS} data exclude transiting short-period planets down to about half an Earth radius. We compare the planetary system architecture of very low-mass stars. In the currently known configuration, the planetary system of Teegarden's star is dynamically quite different from that of TRAPPIST-1, which is more compact, but dynamically similar to others such as GJ\,1002.}


\date{Received 21 September 2023 / Accepted 15 January 2024}

\maketitle
\sloppy

\section{Introduction}

The successful search for exoplanets has been a remarkable achievement in modern astronomy, resulting in the discovery of over 5000 planets outside the solar system.
Furthermore, advancements in radial velocity (RV) surveys have led to an increasing number of low-mass planet detections.
A small subset of these known planets holds special interest: rocky planets located within the habitable zone of their host star, where conditions could potentially support liquid water on their surfaces \citep{1993Icar..101..108K,2013ApJ...765..131K}. The Habitable Exoplanets Catalog currently lists only 24 Earth-sized planets in the conservative sample of potentially habitable exoplanets\footnote{\url{https://phl.upr.edu/projects/habitable-exoplanets-catalog}}. It is noteworthy that the majority of these planets have been found orbiting M dwarfs. Among them, the TRAPPIST-1\footnote{\href{https://www.trappist.uliege.be/cms/c_5006023/en/trappist}{TRAnsiting Planets and PlanetesImals Small Telescope}} system \citep{2017Natur.542..456G}, with four planets in the list, is the only system with precisely determined planetary masses and radii.
Without the requirement of being within the habitable zone, the catalog of Transiting M dwarf Planets (TMP)\footnote{\url{https://carmenes.caha.es/ext/tmp/}} lists 21 potentially Earth-like planets (M$_{\rm planet} < 2 $\,M$_{\oplus}$) including the seven rocky planets of the TRAPPIST-1 system with mass determinations from RVs or transit timing variations.

Multi-planet systems are also of great interest as they provide valuable insights into planet formation and evolution. As of April 2023, a total of 850 multiple planet systems have been discovered\footnote{\url{http://exoplanet.eu/catalog}}. \citet{2021ApJ...907...81L} found that the number of planets in these systems is related to the size of the protoplanetary disk. Additionally, these authors show that the timescale for protoplanet appearance plays a crucial role in determining the planet configuration arising from resonant trapping. A shorter timescale results in a larger number of formed planets, which become trapped in more closely spaced resonances. Their simulations also indicate that resonant planets are generally not formed around stars with masses larger than about 0.4\,M$_\odot$. For a comparison of these predictions with observations, we should therefore aim for the most complete knowledge of the number of planets in multi-planet systems. Another important aspect seems to be the description of the gas disk interaction with the planetary embryos, as discussed by \citet{2022A&A...663A..20S}. The similarity in planet properties and the arrangement of their orbits, forming a chain of mutual resonant configurations, led \citet{2017A&A...604A...1O} to conclude that the TRAPPIST-1 system and its planetary architecture can be well explained by the growth of planets driven by pebbles at the water ice line, followed by inward migration. As discussed by \citet{2023ASPC..534..717D}, such planetary systems offer valuable constraints on planet formation scenarios, thereby motivating further characterization efforts.

Earlier studies have indicated that low-mass stars often host Earth mass planets at relatively short orbital periods \citep{2013ApJ...767...95D,2015ApJ...807...45D}. In such systems, planets within the habitable zone are located closer to the star, making them more accessible for investigations. Moreover, M dwarfs, which represent the most common class of stars in the solar neighborhood \citep[e.g.,][]{2021A&A...650A.201R, 2023A&A...670A..19G}, have extremely long-lived main sequence phases, making them excellent candidates for studying potentially habitable systems. Planets around nearby M dwarfs are also the most suitable targets for future very high contrast and spacial resolution imaging, allowing to probe exoplanet atmospheres. The planetary atmosphere investigations will be possible with  ground-based instruments in the era of extremely large telescopes in reflected light with instruments like the ArmazoNes high Dispersion Echelle Spectrograph \citep{2023arXiv231117075P} or the Planetary Camera Spectrograph \citep{2021Msngr.182...38K} and with space missions such as the {\em Habitable Worlds Observatory} ({\em HWO}) which has been recommended to NASA as the next flagship mission by the US Astro 2020 Decadal Survey report  \citep{NAP26141}
or the complementary {\em Large Interferometer For Exoplanets (LIFE)} mission \citep{2022ExA....54.1197Q,2022A&A...664A..21Q} in thermal emission. \citet{2023A&A...678A..96C} investigated the detectability of exoplanet atmospheres around stars within 20\,pc. From the 212 planets detectable (signal-to-noise $>7$ in less than 100\,h) with the reference configuration of {\em LIFE}, 49 can also be detected with the notional {\em HWO}, 163 with {\em LIFE} only. From the 38 {\em LIFE}-detectable planets in the habitable zone, 13 are below five Earth masses.

\begin{table*}
    \centering
    \caption{List of RV datasets$^{(a)}$.}
    \begin{tabular}{@{}lllccccc@{}}
        \hline
        \hline
        \noalign{\smallskip}
    Instrument & Acronym    & Baseline & \multicolumn{3}{c}{Measurements} & Int. prec. & rms \\
               &            &          & o & c & b & (m/s) & (m/s) \\
        \hline
        \noalign{\smallskip}
    CARMENES & CARM~VIS         & January 2017--March 2023   & 262 & 250 & 230 & 1.67 & 1.73\\
    ESPRESSO & ESPRESSO             & September 2019             & 11 & 11 & 5     & 0.69 & 0.37\\
    MAROON-X Red 1 & MX R1 & August 2021       & 9 & 9 & 9     & 0.32 & 0.67\\
    MAROON-X Blue 1 & MX B1 &                   &  & &           & 1.59 & 1.88\\
    MAROON-X Red 2 & MX R2 & October 2021      & 7 & 7 & 7     & 0.35 & 0.92\\
    MAROON-X Blue 2 & MX B2 &                   &  & &           & 1.57 & 1.98\\
    MAROON-X Red 3 & MX R3 & August 2022       & 8 & 8 & 7     & 0.25 & 1.11\\
    MAROON-X Blue 3 & MX B3 &                   &  & &           & 0.97 & 0.80\\
    HPF & HPF & September 2019--October 2021      & 146 & 145 & 74 & 2.26 & 2.64\\
        \noalign{\smallskip}
    \hline
        \noalign{\smallskip}
    All & & & 467 & 454 & 355 &\\
        \noalign{\smallskip}
        \hline 
   \end{tabular}
\tablefoot{
\tablefoottext{a}{For the number of measurements those of the original (o), the 5\,$\sigma$ clipped (c), and daily binned data (b) sets are listed. The rms was calculated using the residuals after subtracting the best-fit model. The internal precision (the median of the uncertainties) was measured after the binning.}
}
    \label{tab:inst}
\end{table*}

One particularly remarkable system is Teegarden's Star, an M7.0 dwarf \citep{2015A&A...577A.128A} discovered by \citet{2003ApJ...589L..51T}. This is the 25$^{\mathrm{th}}$ nearest star to the Sun\footnote{\url{https://gruze.org/10pc/}} \citep{2021A&A...650A.201R}, at a distance of only 3.831\,pc. Another significant attribute of Teegarden's Star is its relatively low magnetic activity compared to most late-M dwarfs, which enhances its potential as a target for the search for extraterrestrial life. In 2019, \citet{2019A&A...627A..49Z} presented evidence of two Earth mass planets within the potentially habitable zone of Teegarden's Star, with orbital periods of 4.91 and 11.4\,days, respectively. Both planets are among the 13 low-mass habitable-zone planets detectable with {\em LIFE}. These were the first, and at present still are, the only planets detected around an ultra-cool dwarf (spectral type later than M\,7.0\,V) using radial velocities. Subsequent data collection allows us now to search for weaker signals in the system so that a more complete inventory of low-mass planets can be obtained, providing more reliable constraints on planetary architectures and properties around very low-mass stars. Here we report the discovery of a third planet orbiting Teegarden's Star and find evidence for further suggestive signals that could point to an even larger number of planets in the system, thereby resembling TRAPPIST-1. 

First, we introduce the RV measurements, the spectroscopic activity indices, and transit data in Sect. \ref{sect:data}. In Sect. \ref{sect:methods}, we describe the methods used for analysis. The stellar parameters and results are presented and discussed in Sect. \ref{sect:results} and Sect. \ref{sect:discussion}, respectively.

\section{Observations and data products}
\label{sect:data}

\subsection{Radial velocity data}

RV measurements were obtained with four different instruments described below. Based on the instrument-specific data reduction we computed the RVs from all instruments with {\tt serval}\footnote{\url{https://github.com/mzechmeister/serval}} \citep{2018A&A...609A..12Z}.

The CARMENES instrument (Calar Alto high-Resolution search for M dwarfs with Exoearths with Near-infrared and optical Echelle Spectrographs) at the 3.5\,m telescope at the Calar
Alto Observatory in Almería, Spain, is a dual-channel spectrograph that operates at both optical (VIS, 0.52--0.96\,$\mu$m) and
near-infrared (NIR, 0.96--1.71\,$\mu$m) wavelengths. 
The average resolving power for the two wavelength regions is $\mathcal{R} = 94\,600$ and
$\mathcal{R} = 80\,400$, respectively. 
We used the available 262 measurements (July 2023) of the RV 
of Teegarden's Star from the guaranteed time (GTO) and legacy project 
observations of the CARMENES project \citep{2014SPIE.9147E..1FQ,2023A&A...670A.139R}. Compared to the 239 measurements with nightly zero point corrections \citep{2018A&A...609A.117T} used by \citet{2019A&A...627A..49Z}, we
now added 14 more published by \citet{2023A&A...670A.139R}. Additionally to the GTO data, nine measurements were taken in the last observing season (July 2022 to March 2023) as part of the CARMENES Legacy-Plus project. Due to the larger scatter of the NIR data and the small signals expected, we restricted the analysis to the VIS data of CARMENES. \cite{2020A&A...640A..50B} compared the scatters of the VIS and NIR RV curves for an M dwarf of roughly the same $J$ magnitude as Teegarden's Star. During part of the time during which these data where acquired, the NIR channel had a peak-to-peak instrumental drift substantially larger than that of the VIS channel, and suffered from instrumental changes during maintenance operations of the active cryogenic cooling system. We like to note that recent instrumental updates have significantly reduced this effect. 
The impact of the telluric lines on the RV measurements has been tested using CARMENES spectra, which were cleaned from telluric contamination \citep{2023arXiv231014715N}, but we found very small differences compared to the standard results from {\tt serval}. Nevertheless, we used the telluric-corrected RV data.

ESPRESSO \citep{2021A&A...645A..96P} is the Echelle SPectrograph for Rocky Exoplanets and Stable Spectroscopic Observations installed at the incoherent combined coudé facility of the Very Large Telescope (VLT). It is an ultra-stable fiber-fed \'echelle high-resolution spectrograph ($\mathcal{R} = 140\,000$,  0.38-–0.79\,$\mu$m in high resolution single unit telescope mode). We obtained 11 ESPRESSO measurements in September 2019. 

MAROON-X \citep{2016SPIE.9908E..18S}
is a stabilized, fiber-fed high-resolution ($\mathcal{R} = 85\,000$) spectrograph mounted at the 8.1\,m Gemini North telescope on Mauna Kea, Hawai'i, USA. It has blue and red arms, which encompass 0.50--0.68\,$\mu$m and 0.65--0.92\,$\mu$m, respectively. During an observation, both arms are operated simultaneously. MAROON-X obtained nine, seven, and eight measurements with the blue and red arms in three observing runs in August and October 2021, and in August 2022, respectively.

The Habitable Zone Planet Finder (HPF) is a stabilized fiber-fed near-infrared spectrograph (0.9--1.8\,$\mu$m) for the 10\,m Hobby-Eberly Telescope \citep[HET,][]{1998SPIE.3352...34R} with a resolution of $\mathcal{R}=53\,000$ \citep{2012SPIE.8446E..1SM}. Between September 2019 and October 2021, a total of 143 measurements were collected, with each night's observations organized in pairs of adjacent measurements.
The spectra were extracted with the HPF pipeline {\tt HxRGproc}  \citep{2018SPIE10709E..2UN, 2019ASPC..523..567K, 2019Optic...6..233M}.

Overall, we have 228 new measurements extending the time baseline by about three years compared to \citet{2019A&A...627A..49Z}.
In case of multiple observations per night, we used nightly averages with the nightly standard deviation as uncertainty for the night. We also excluded outliers using a $5\,\sigma$-clipping procedure. This resulted in 355 measurements after $\sigma$-clipping and nightly binning for the full dataset (Table~\ref{tab:inst}). The original RV data, that were un-binned and un-clipped, are available at CDS. A short section is shown in Table~\ref{tab:RVs}.

\subsection{Spectroscopic indices}\label{sect:specindex}

From the CARMENES and MAROON-X data, we extracted several spectroscopic indices using {\tt serval}. Here we make use of the chromatic index (CRX), the differential line width (dLW), and the emission strength of the hydrogen H$\alpha$ line. In short, the CRX index measures the wavelength dependence of the RVs determined in each spectral order. While planets do not have a color-dependent RV, stellar activity signals from spots and plagues are likely to show them due to the surface temperature modulations. The dLW is similar to the full width half maximum (FWHM) index from cross-correlation techniques. It measures line profile variations, which can either originate from instrument changes or from stellar activity. The equivalent width of the hydrogen H$\alpha$ line is notoriously difficult to determine in M dwarfs due to the absence of a continuum free of spectral lines. A slowly rotating template stars for each M spectral subclass is therefor use as reference. For details about the indices we refer to \citet{2018A&A...609A..12Z} and \citet{2019A&A...623A..44S}.  

Although the spectra were cleaned from telluric lines \citep{2023arXiv231014715N} and wavelength regions with strong telluric contamination were masked for RV measurements, residual telluric lines may still have an impact. We simulated the impact of telluric lines on the RV determination using an appropriate synthetic spectrum and adding a telluric spectrum shifted by the barycentric velocity of each observing date. No planetary signal was added. The result was a time series of simulated spectra, which were then analyzed using {\tt serval} to determine the RVs. Without a planetary RV signal in the simulated data a measured RV is then due to telluric lines, resulting in a telluric contamination index. 

\subsection{Photometric data}

The {\em Transiting Exoplanet Survey Satellite} \citep[{\em TESS};][]{Ricker2014SPIE, 2015JATIS...1a4003R} observed Teegarden's Star in sectors 42 and 44 in full frame image mode and in sectors 43, 70, and 71 with 120\,s cadence. The uncertainty of the normalized flux in this light curve is 0.002, which motivated the search for transit signals of small planets (see Sect.\,\ref{sect:transits}). 

As a nearby ultracool dwarf, Teegarden's Star has been observed by the SPECULOOS (Search for habitable Planets EClipsing ULtra-cOOl Stars) project, which aims at finding transiting planets around such stars \citep[][]{2018haex.bookE.130B,2018SPIE10700E..1ID,2021A&A...645A.100S}. In particular, Teegarden's Star was observed by SPECULOOS telescopes located in the northern hemisphere, namely Artemis \citep{2022PASP..134j5001B} and SAINT-EX \citep{2020A&A...642A..49D}, which both can reach photometric precisions for a few minutes sampling of $\sim$0.1\% and mid-transit times accuracies of $\sim$1\,min \citep[see e.g.,]{2022A&A...667A..59D,2023A&A...672A..70P}. In total, Teegarden's Star was observed for 142\,h over 35 nights between August 2021 and December 2022 (see Sect.\,\ref{sect:speculoos}).

\section{Methods}
\label{sect:methods}

The RVs were analyzed using a multi-planetary model employing Keplerian orbits. The Keplerian model was parameterized with the semi-amplitude $K$, orbital period $P$, $h=\sqrt e\sin\omega$ and $k=\sqrt e\cos\omega$ instead of eccentricity $e$ and the argument of periastron $\omega$, and the mean longitude $\lambda$, while the inclination $i$ and the longitude of the ascending node $\Omega$ remain undetermined in the Keplerian model. Without limiting the generality, we therefore set $i=90^\circ$ and $\Omega=0^\circ$. This model was implemented in a series of scripts in python language. 

As a cross-check, we used the versatile modeling tool {\tt juliet} \citep{2019MNRAS.490.2262E}, which implements RV fits using {\tt RadVel} \citep{2018PASP..130d4504F}, a python package for modeling Keplerian orbits in RV time series. In addition to the aforementioned parameters ($K$, $P$, $h$, and $k$), the Keplerian model in {\tt juliet} was parameterized with the time of periastron passage $T$, since {\tt juliet} currently does not support the mean longitude $\lambda$ as a parameter. We compared the results and the estimated values of the logarithm of the Bayesian evidence ($\ln \mathcal{Z}$) between the models calculated with {\tt juliet} and those previously described in order to ensure that the results were consistent. We also used {\tt Exo-Striker} \citep{2019ascl.soft06004T}, which allows efficient testing and visualization of modeling approaches. 

We complemented the planetary models of varying complexity with a Gaussian process (GP) approach to model possible contributions from stellar activity, particularly in the form of rotational modulation. The posterior distributions of the parameters and hyper-parameters were determined with a nested sampling algorithm, which also provides the Bayesian evidence that was used to identify the best model. Finally, the parameters of the best model were checked for orbital stability following a dynamical analysis.

In an alternative approach, we used the $\ell_1$-periodogram\footnote{\url{https://github.com/nathanchara/l1periodogram/tree/master}} \citep{2017MNRAS.464.1220H} to identify significant signals in the RV data. Based on an input frequency grid and a model for the covariance of the data, the $\ell_1$-periodogram searches for a representation of the data with a small number of sinusoidal signals. The algorithm simultaneously tests all frequencies of the input grid. Following the tutorial from the {\tt git} repository, we also tested the robustness of the noise model. A grid with various amplitudes for white noise, red noise, and correlated noise as well as periods and time scales for red noise and correlated noise provided a cross validation score for each covariance matrix.

\subsection{Gaussian process}
\label{sect:GP}

GP regression is a nonparametric Bayesian method used for modeling complex functions by assuming that the underlying function is a sample from a Gaussian distribution. The python package {\tt celerite} \citep{2017AJ....154..220F} provides an efficient implementation of GP regression, specifically designed for large datasets, by modeling the covariance matrix as a sum of complex exponential functions. For modeling stellar activity, a simple damped harmonic oscillator (SHO) kernel is suitable 
(details are given in Equations 9, 20, and 21 of \citealt{2017AJ....154..220F})
We use an improved kernel for rotation-modulated stellar activity composed of two SHO kernels (here called dSHO) from \citet{2018RNAAS...2...31F}. The hyper-parameters for both SHO kernels include the scale of the activity induced noise $\sigma_0$ for each instrument, the quality factor of the oscillator $Q_0$, the scale ratio between the two oscillators $f$, the difference of the quality factors $dQ$, and the period of the underlying process $P$. The period of the second oscillator was fixed to the second harmonic of the first in order to also account for stellar activity signals at half of the rotation period. Additional white noise was incorporated using a jitter term. Following the discussion of \citet{2023AJ....166...62B} about potential overfitting with GP kernels of too much flexibility, we also tested a dSHO kernel, where the damping time scale of the two oscillators is forced to be identical. We also tested the effect of a joint covariance matrix for all RV datasets using their modified version of {\tt RadVel}. The impact on the planet parameters of both is negligible. This is not unexpected since the correlated noise impact is small (see below).

\subsection{Nested sampling}

Nested sampling is a Bayesian inference method for estimating the logarithm of the evidence ($\ln \mathcal{Z}$) and the posterior distributions of a model. The python package {\tt dynesty} \citep{2020MNRAS.493.3132S} is an implementation of nested sampling that is efficient in exploring complex posterior distributions and has a dynamic nested sampling algorithm. It provides diagnostic tools and options for controlling the sampling process. We used 5000 live points and stopped when $\delta \ln \mathcal{Z}<0.01$. We used the {\it multi} mode and the {\it rwalk} sampling. The priors for all of the parameters are listed in Tables\,\ref{tab:modresult_p} and\,\ref{tab:modresult_d}. 

\begin{figure*}
    \centering
    \includegraphics[width=0.99\textwidth]{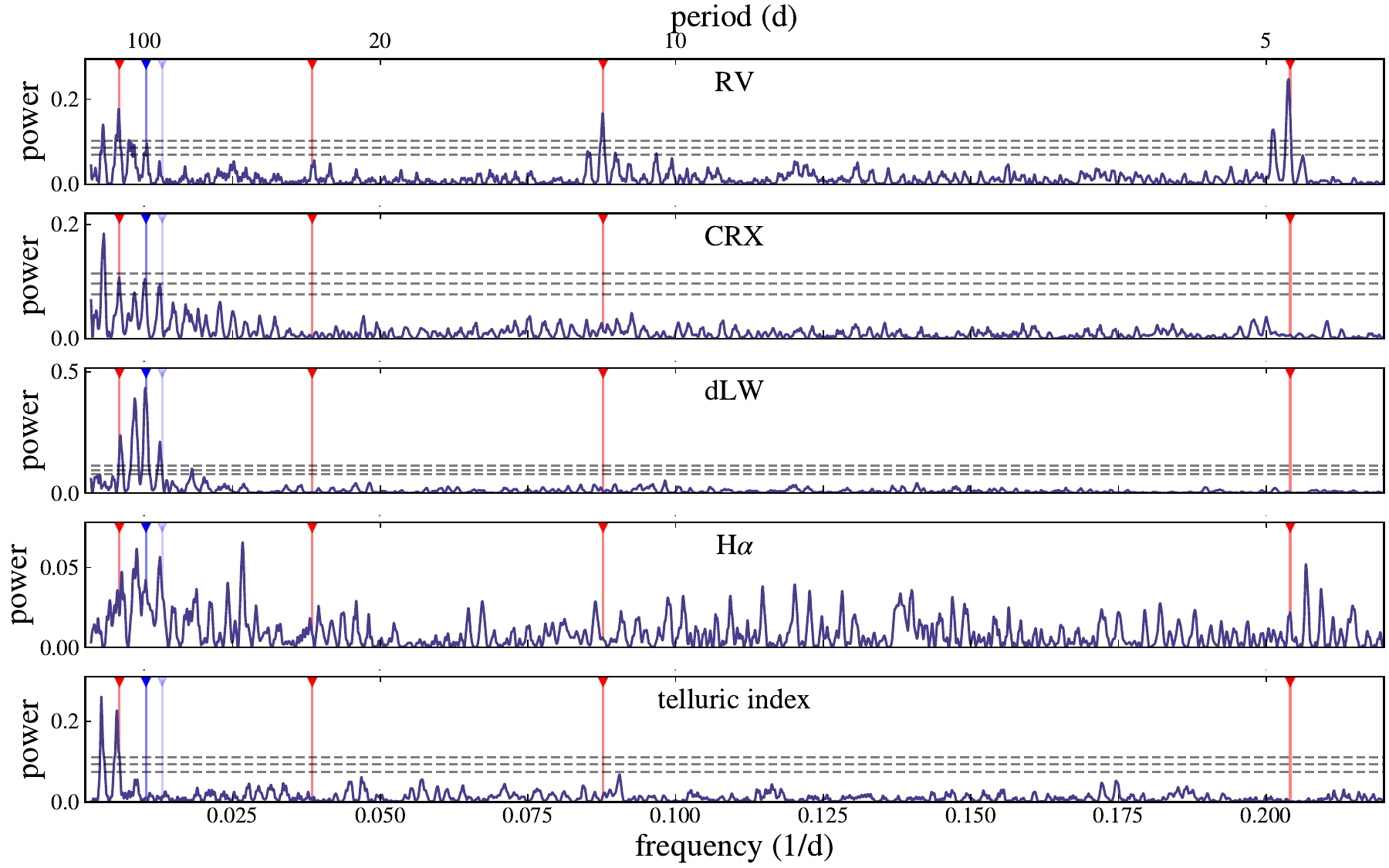}
    \caption{Periodogram of RV data and spectroscopic activity indicators from CARMENES VIS. From top to bottom: RV data, chromatic index (CRX), differential line width (dLW), H$\alpha$, and the telluric contamination index. 
    The blue triangles indicate the rotation period at 96\,days and its 1-year alias at 79\,days, the red triangles indicate the periods of the other signals (4.9\,days, 11.4\,days, 26\,days, and 172\,days). The dashed horizontal lines show the false alarm probability levels at 10\%, 1\% and 0.1\%. }
    \label{fig:activity}
\end{figure*}

\subsection{\tt SPOCK}
\label{sect:SPOCK}

 {\tt SPOCK} \citep[Stability of Planetary Orbital Configurations Klassifier;][]{2020PNAS..11718194T} is a machine learning algorithm that was designed to predict the long-term stability of planetary systems. The algorithm is trained on a set of numerical simulations of planetary systems and uses a combination of physical and dynamical features to classify the stability of a given system. {\tt SPOCK} has been used to predict the stability of observed exoplanetary systems \citep[e.g.,][]{2020PNAS..11718194T,2021MNRAS.501.4798T,2022MNRAS.510.5464K}, and has been shown to be highly accurate in identifying stable systems. Additionally, it has been used to identify new configurations of exoplanetary systems that are likely to be stable, which can help guide the search for and validation of new exoplanets \citep{2023arXiv230204922W}. We used {\tt SPOCK} to check the dynamical stability of the orbital solutions (see Sect.\,\ref{sect:stability}).

\section{Results}
\label{sect:results}

\subsection{Stellar properties}
\label{sect:stellar}

The stellar atmospheric parameters ($T_{\rm eff}$, $\log{g}$, and [Fe/H]) of Teegarden's Star in Table~\ref{tab:stepar} were derived from the co-added CARMENES spectra with the {\sc SteParSyn}\footnote{\url{https://github.com/hmtabernero/SteParSyn}} code \citep{2022A&A...657A..66T} using the line list and model grid described by \citet{2021A&A...656A.162M}. 
This update led to small modifications compared to the values reported by \citet{2019A&A...627A..49Z}. All planetary masses listed in this paper are derived relative to the updated stellar mass of Teegarden's Star, and the uncertainties consider the error propagation from the stellar mass. 

Based on spectroscopic indices (see Sect.\,\ref{sect:specindex}), the rotation period of Teegarden's Star was determined as 96\,d by \citet{2021A&A...652A..28L}. A similar period of 100\,d was found in the HPF data by \cite{Terrien2022ApJ...927L..11T}. A cluster analysis of all spectral activity indicators revealed 98\,d as the most likely stellar rotation period \citep{kemmer}, aligning well with the determinations made by \citet{2021A&A...652A..28L} and \citet{Terrien2022ApJ...927L..11T} using spectroscopic activity indicators and Zeeman signatures, respectively. This agrees with earlier estimates of the rotation period reported by \citet{2019A&A...627A..49Z} using results from \citet{2015ApJ...812....3W}, \citet{2017ApJ...834...85N}, and \citet{2018A&A...614A..76J}. The long rotation period also matches well with the low stellar activity indicating an age of around 8\,Gyr as discussed in more detail by \citet{2019A&A...627A..49Z}. 

In Fig.\,\ref{fig:activity}, we present the periodograms of the CARMENES RV data, along with selected spectroscopic indices. Notably, the potential rotation period at 96\,d and its 1-year alias at 79\,d are evident in the RVs (first panel), the CRX (second panel), and dLW (third panel) indices. However, the spectroscopic indices and the RVs share more long-periods (320\,d, 172\,d, 120\,d), with the most prominent one among them in the RV data occurring at 172\,d.

The power observed at 320\,d and 120\,d can be attributed to spectral leakage from the window function (see the bottom panel in Fig.\,\ref{fig:simulation}). This spectral leakage also contributes weakly to the power at 96\,d. Consequently, we may interpret the 172\,d signal as the true rotation period, with the other long-period variabilities being considered as alias signals. However, such a long rotation period would make Teegarden's Star an outlier in terms of slow rotation. \citet{2018AJ....156..217N} list 281 M stars with rotation periods determined from photometric monitoring using MEarth South \citep{2015csss...18..767I}, including 31 stars below 0.12\,M$_\odot$. Only one of them has a rotation period longer than 172\,d, five have rotation periods above 150\,d \citep[see Fig.\,4][]{2018AJ....156..217N}. A similar result is presented by \citet{2024arXiv240109550S}. In their literature compilation complemented with 129 new determinations using photometric monitoring within the CARMENES project, there is only one star with comparably low mass and a rotation period of 178\,d, two in the range of 150\,d, all other 259 M stars are reported with lower rotation periods. Therefore we conducted further investigations into the signal at 172\,d.

\begin{table}
    \centering
    \caption{Updated stellar parameters compared to those from \citet[][Zec19]{2019A&A...627A..49Z}.}
    \begin{tabular}{@{}lccr@{}}
        \hline
        \hline
        \noalign{\smallskip}
        Parameter & Zec19 & This work & reference \\
        \noalign{\smallskip}
        \hline
        \noalign{\smallskip}
        $T_{\rm eff}$ [K]& $2904\pm 51$ & $3034\pm45$ & Marf21\\
        $\log g/$[cm/s]  & ... & $5.19\pm0.2$ & Marf21\\
        $[$Fe/H$]$ & $-0.19\pm0.16$ & $-0.11\pm0.28$ & Marf21\\
        $M$ [$M_\odot$]    & $0.089\pm0.009$ & $0.097\pm0.010$ & This work \\
        $R$ [$R_\odot$]    & $0.107\pm0.004$ & $0.120\pm0.012$ & This work \\
        $L$ [$10^{-5}\,L_\odot$]    & $73\pm1$ &$72.2\pm0.5$ & This work \\
        $P_{\rm rot}$ [d] & ... & 96.2 & Laf21\\
        \hline
    \end{tabular}
    \label{tab:stepar}
    \tablebib{Marf21: \citet{2021A&A...656A.162M}, Laf21: \citet{2021A&A...652A..28L}.}
\end{table}

First, we fit Keplerian orbits to the RV data and to the spectroscopic indices using a uniform prior for the orbital period in the range of 150\,d to 190\,d. The result is shown in Fig.\,\ref{fig:hist_indices}. 
In case that residual telluric contamination were the cause of the 172\,d signal in the RV data, we would expect that the period and phase of the RV data and the telluric contamination index agree, which is not the case. As a result, 
we excluded contamination from residual telluric lines. Since the CARMENES data cover about 14 cycles of the 172\,d period, the difference in period and phase seems to be robust. 
The results also indicate that the signal might be related to stellar activity, since the dLW and CRX indices show signals at consistent periods (Fig.\,\ref{fig:hist_indices}). We therefore calculated the stacked Bayesian Lomb-Scargle periodogram \citep[sBGLS;][]{2017A&A...601A.110M} of the observed RVs, displayed in Fig.\,\ref{fig:sbgls} (top). The variability of the power could be an indication of non-coherence and, hence, possibly interpreted as caused by stellar activity. In order to check the impact of the irregular sampling and spectral leakage, we injected a coherent signal corresponding to a Keplerian orbit of 172\,d using the parameters from Table\,\ref{tab:modresult_c} into the residuals of 
our overall best-fit model (model E, see Table\,\ref{tab:modselect}). This synthetic dataset represents a single coherent signal at 172\,d with the noise and sampling characteristics of the observation. The sBGLS periodogram is shown in the bottom panel of 
Fig.\,\ref{fig:sbgls}. The power of the 172\,d signal is also variable, despite being coherent. Since both sBGLS periodograms show variable power at 172\,d in the observed and simulated data, the power fluctuations are therefore attributed to variable spectral leakage rather than to non-coherence. The 172\,d signal is therefore compatible with being coherent over 2\,500\,d. It remains unclear whether and according to what mechanism the signal at 172\,d is related to stellar activity or other subtle effects such as remaining sky emission or detector artifacts. In this study we keep 96\,d as the rotation period and discuss a potential planetary origin for the 172\,d signal in the following section.

Inspection of the {\it TESS} data does not provide information about the stellar rotation period. Using the Simple Aperture Photometry (SAP) data we cut out regions with obvious excess instrumental noise. The light curve of the sectors 43 and 70 each show a small trend of about 0.01 flux change over the sector length, in sector\,43 it is a positive, in sector\,70 a negative trend. No trend is visible in sector 71. The sectors are separated by two years the gap is therefore too large to obtain a meaningful fit for a periodic variation. Within each sector, variability at the level of the rms on time scales below 10\,d can be seen. Sector 43 additionally shows two small flares. The {\it TESS} light curves presents Teegarden's Star as a photometrically very quite star. This well matches with the conclusion from \citet{2019A&A...627A..49Z}, where occasional flares as well as an overall low activity level were concluded from H$\alpha$-variability.

The spectroscopic activity indices as well as the RV data show a trend at a timescale of the length of the dataset. Fitting a sinusoidal function reveals a period of at least the length of the dataset ($\sim 2\,500$\,d, Fig.\,\ref{fig:hist_cycle}). Since the posterior distributions of the period and phase overlap, this may constitute an indication of an activity cycle of unknown duration.

\subsection{Analysis of CARMENES data}
\label{sect:CARMENES}

The measurements from CARMENES VIS are the ones with the longest time baseline, in addition to having a quite dense sampling and relatively low uncertainties. We therefore first analyzed the CARMENES VIS data separately and later turned to the analysis of the combined dataset. 

\label{sect:select}

\begin{figure}[ht]
    \centering
    \includegraphics[width=0.49\textwidth]{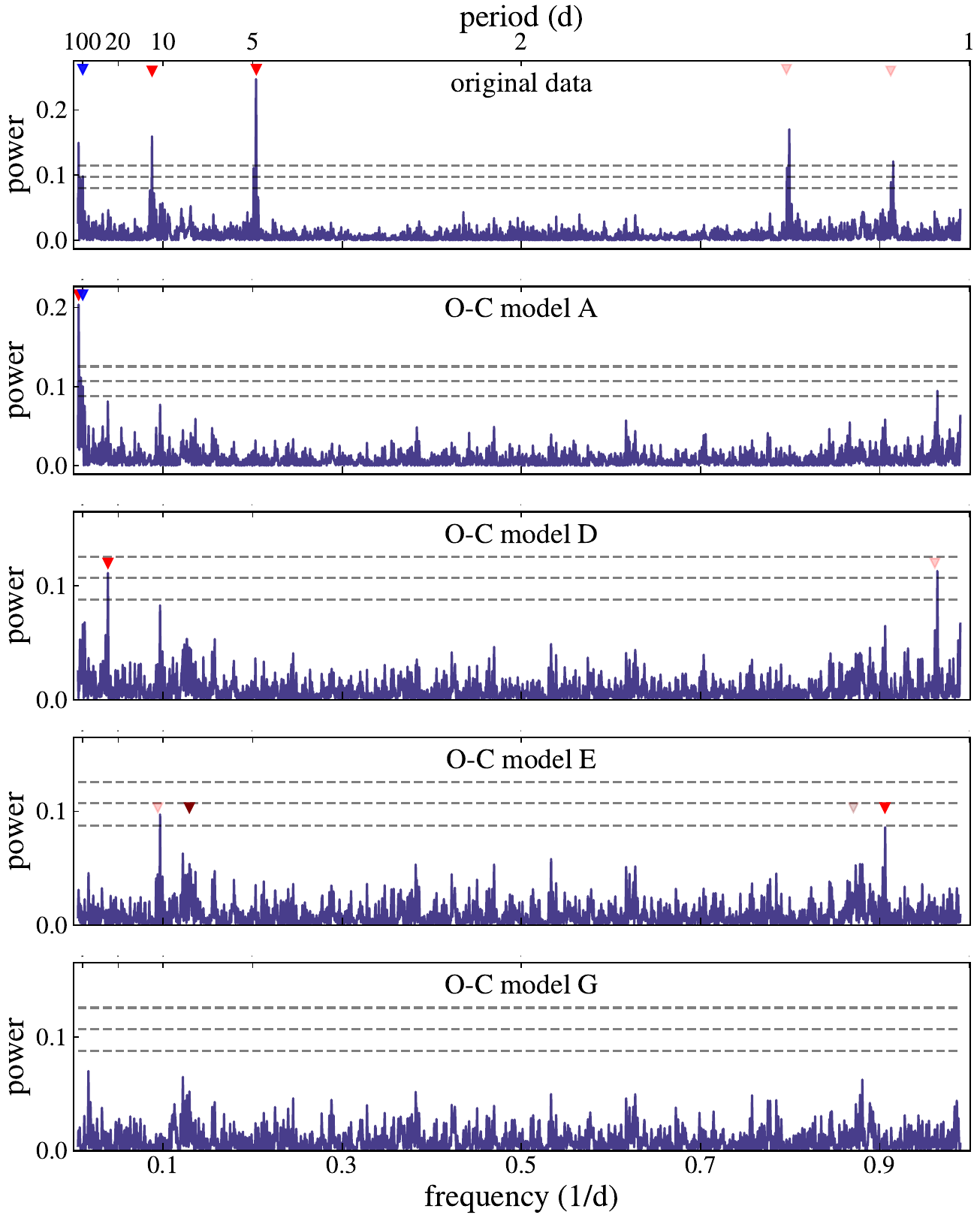}
    \caption{Periodogram of the CARMENES VIS data. From top to bottom, signals are subsequently subtracted, with red color denoting potential planetary signals and blue indicating the rotation period. Model names correspond to those in Table\,\ref{tab:modselect}. The 1\,d alias period of the signals is marked with a triangle in lighter color. The dashed horizontal lines are the false alarm probabilities at 10\%, 1\% and 0.1\%. From top to bottom: Original data, model A, model D, model E, and model G.}
    \label{fig:periodogram}
\end{figure}

\begin{table}[t]
    \centering
    \caption{Model selection based on Bayesian evidence$^{(a)}$.}
    \begin{tabular}{llllllr}
        \hline
        \hline
        \noalign{\smallskip}
        \multicolumn{2}{c}{Model} & \# & GP & $\ln \mathcal{Z}_{\rm CARM}$ & $\ln \mathcal{Z}_{\rm Full}$& $\Delta \ln \mathcal{Z}_{\rm Full} $\\ 
        \noalign{\smallskip}
        \hline
        \noalign{\smallskip}
        A & b and c     & 2 & no & $-$539.8 & $-$845.9 & $-$29.0 \\ 
        B & A+172\,d    & 3 & no & $-$522.5 & $-$830.4 & $-$13.5 \\ 
        {\bf C} & {\bf A+26\,d} & 3 & no & $-$536.0 & $-$829.4 & $-$12.5 \\ 
        D & B+96\,d     & 4 & no & $-$517.9 & $-$825.4 & $-$8.5 \\
        {\bf E} & D+26\,d & 5 & no &  $-$510.3 & {\bf $-$816.9} & {\bf 0.0} \\
        F & E+7.7\,d    & 6 & no & {\bf $-$507.8} & $-$815.8 & +1.1 \\
        G & E+1.1\,d    & 6 & no & $-$508.7 & $-$814.9 & +2.0 \\
        \hline
        H &             & 2 & yes& $-$523.4 & $-$835.2 & $-$18.3\\
        I & H+26\,d     & 3 & yes& $-$515.3 & $-$827.8 & $-$10.9\\
        J & I+172\,d    & 4 & yes& $-$513.2 & $-$826.0 & $-$9.1\\
        \noalign{\smallskip}
        \hline
     \end{tabular}
\tablefoot{
\tablefoottext{a}{Columns denote the model name, a short description, the Bayesian evidence for the model using CARMENES VIS data only as well as for the full dataset, and the difference with respect to our best model.}
}

     \label{tab:modselect}
\end{table}

The periodogram of the RV data (Fig.\,\ref{fig:periodogram}, top panel) shows significant peaks from the two already reported planets, namely Teegarden's Star b and c \citep{2019A&A...627A..49Z}. By subtracting these two signals (model A, see Table\,\ref{tab:modselect}), the periodogram of the residuals shows the strongest power at 172\,d (Fig.\,\ref{fig:periodogram}, second panel). The inspection of the periodogram of the spectroscopic activity indicators (Fig.\,\ref{fig:activity}, see Sect.\,\ref{sect:stellar}) reveals that this period is present there as well. Nevertheless, we fit it with a Keplerian orbit (model B), which then leads to the next highest peak at 96\,d, which we relate to stellar rotation (Sect.\,\ref{sect:stellar}). 
Subtracting this signal (model D) results in a peak at 26\,d (Fig.\,\ref{fig:periodogram}, third panel). We interpret this as the signal of a third planet, Teegarden's Star\,d. The Bayesian evidence $\ln \mathcal{Z}$ of model E increases by more than 5, which indicates that the more complex model is superior (Table\,\ref{tab:modselect}, fourth column). We also tested whether adding the 26\,d signal to model A (model C) results in an improved Bayesian evidence. The difference of $\Delta \ln \mathcal{Z} = 3.8$ indicates a moderate improvement.

We checked whether the five signals exhibit any crosstalk due to spectral leakage in the window function of the measurement time series. The potential spectral leakage can be investigated in Fig.\,\ref{fig:simulation}.
The three planetary signals at 4.9\,d, 11.4\,d, and 26\,d are not affected by crosstalk with any other signals. We note that the 11.4\,d signal produces some power close to, but not exactly at, 26\,d. Crosstalk between the two signals is therefore not expected. It should be noted that the 172\,d signal has a side lobe very close to 96\,d. However, the removal of the 172\,d signal eliminated the aliases at 320\,d and 120\,d but did not affect the signal at 96\,d. As a result, these two periodicities at 96\,d and the 172\,d seem to be independent but possibly affected by spectral leakage. 

After subtracting the five signals, no remaining peak reaches a power above the 10\% false alarm probability level in the periodogram, but two signals appear to be stronger than the adjacent noise. The one at 7.7\,days is especially suggestive because it lies close to a 3:2 period commensurability chain with the two known planets (at 4.9\,d and 11.4\,d). Again, the Bayesian evidence grows (model F), at least by more than 2.5, so that the model F including an additional planet candidate at 7.7\,days is moderately preferred. 

There are two more peaks of similar strength in the periodogram, one at 1.104\,d, the other its one-day alias at 10.6\,d. If we associate the signal to a planet, an orbital period of 10.6\,d can be ruled out since this would make the system dynamically unstable due to the close vicinity of planet c at 11.4\,d. The 1.104\,days peak would therefore correspond to the true signal if it were a planet. Including a Keplerian at 1.104\,d (model G) instead of the 7.7\,d (model F) would also moderately increase the Bayesian evidence compared to model E. The difference between the Bayesian evidences of the six-signal models is insignificant, but it is tempting to accept the 7.7\,d candidate signal as real since it would make the Teegarden's Star planetary system dynamically compact and remarkably similar to the TRAPPIST-1 system. 

In addition to fitting the potentially activity related signals above $\sim\,70$\,d with Keplerian orbits, we further tried to use the GP dSHO kernel described in Sect.\,\ref{sect:GP} to model the variability in that range. The Bayesian evidence difference between the two-planet model with GP (model H) and the corresponding four-signal model without GP (model D) as well as the three-planet model with GP (model I) compared to the five-signal model without GP (model E) results in a higher evidence for the models without GP, which would support the interpretation of all five signals as being associated with planetary orbits. However, due to the presence of the 96\,d and 172\,d signals in the activity indices (Sect.\,\ref{sect:stellar}), we prefer not to make a claim for the presence of further planets in the system at this point. The signals potentially caused by stellar activity are better represented by variability coherent at least over the duration of the observations indicating a very stable activity pattern on the stellar surface. We also tested a model with four planets and stellar rotation at 96\,d modeled with a dSHO kernel (model J). This is superior compared to one where all long-period signals are modeled by the dSHO kernel (model I). Using CARMENES data alone, model J is moderately preferred. It is worthwhile to mention that the resulting amplitudes of the GP (dSHO kernel with and without enforced identical damping time scales as well as a single oscillator SHO kernel) in model I and J are very close to the amplitudes of the corresponding Kepler models for the 172\,d (1.3\,m\,s$^{-1}$) and 96\,d (1.0\,m\,s$^{-1}$) signals, respectively. The high $Q$ value corresponds to a correlation time scale of about 15\,000\,d, which is larger that the length of the observations. The SHO or dSHO kernels therefore mimic a coherent Keplerian model.

Using the $\ell_1$-periodogram \citep{2017MNRAS.464.1220H} supports our results from the signal detection using Keplerian models of increasing complexity. The $\ell_1$-periodogram with the logarithm of the Bayes factor is shown in Fig.\,\ref{fig:l1periodogram}. The five signals of model E are also present in the $\ell_1$-periodogram and indicated as significant according to their Bayes factor. The long-period signal interpreted as activity cycle (Sect.\,\ref{sect:stellar}, Fig.\,\ref{fig:hist_cycle}) is also present and significant in the $\ell_1$-periodogram.

\subsection{Analysis of the full dataset}
In a second step of the analysis, we
used the full dataset and reran the models from Sect.\,\ref{sect:CARMENES}. We would like to point out that the datasets from ESPRESSO and MAROON-X contain only a relatively small number of measurements. Since the ESPRESSO data overlap with the CARMENES data, the instrumental offset can be rather well constrained. The MAROON-X data happen to correspond to epochs that fall within a long observing gap in the CARMENES time series. However, the HPF dataset overlaps with the CARMENES observations and with two of the three MAROON-X campaigns. The disadvantage of adding multiple short duration datasets, each with an individual instrumental offset, is therefore mitigated by mutual overlaps between the datasets except for the third campaign of MAROON-X. 

\begin{figure}
    \centering
    \includegraphics[width=0.49\textwidth]{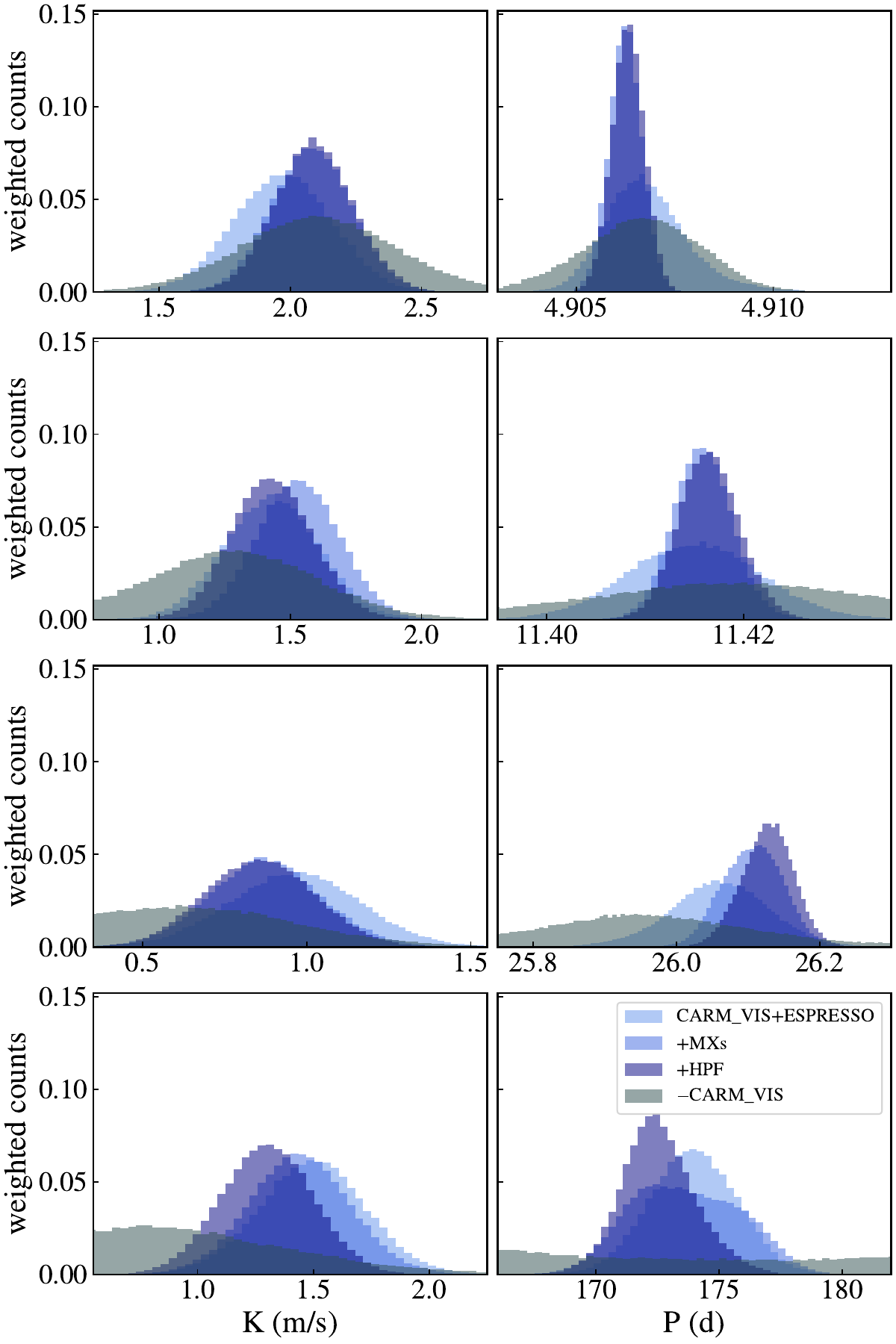}
    \caption{Comparison of the posterior distributions of the three planets (planet b, c, and d from top to bottom) and the signal of an unclear nature at 172\,d depending on the datasets used for the analysis.}
    \label{fig:hist_planets}
\end{figure}

We investigated the impact of the additional data on the planetary parameters. The results are illustrated in Fig.\ref{fig:hist_planets} for the orbital period and RV amplitude. We note the slight improvement of the parameter determination through lower uncertainties (e.g., the orbital period of planet b) and mutual consistency of the parameter determinations within the uncertainties. Without the CARMENES data (gray histogram), the signals of the known planets b and c can be detected, but with significantly larger uncertainties because the time coverage is much smaller. The signal at 26\,d and 172\,d cannot be detected without CARMENES. The ESPRESSO and MAROON-X observations are too short each to cover at least one period, the uncertainty of the mutual offsets therefore interferes with the detection of the signals. The scatter in the HPF data connecting them is too large to compensate that.

In Table\,\ref{tab:modselect} (fifth and sixth columns) the Bayesian evidences and the differences to our preferred model are listed. While the CARMENES data alone would provide statistically significant support for an additional planet between planet b and c at 7.7\,d, the full dataset does not. The short period signal at 1.104\,d is also insignificant in the full dataset. The CARMENES data also gives preference to a model with four planets and stellar rotation at 96\,d modeled as dSHO (model J) compared to a three-planet model and stellar activity variability at periods above $\sim 70$\,d (model I). This, however, is not the case for the full dataset.  

The model containing five Keplerian signals (model E) is our preferred final best model. Three of the signals, those at 4.9\,d, 11.4\,d, and 26\,d, are interpreted as arising from the Keplerian motions of planet b, c, and d in the system, while the signals at 96\,d and 172\,d are interpreted as likely to be caused by stellar activity. Nonetheless, we emphasize that the latter ones are best represented by coherent variability over the entire time baseline of our observations, thus indicating a long-lived stellar activity pattern. The inclusion of the 26\,d signal (planet\,d) leads to an improved Bayesian evidence, independent of the sequence in which signals are added to the models (Table\,\ref{tab:modselect}).

\begin{figure}
    \centering
    \includegraphics[width=0.49\textwidth]{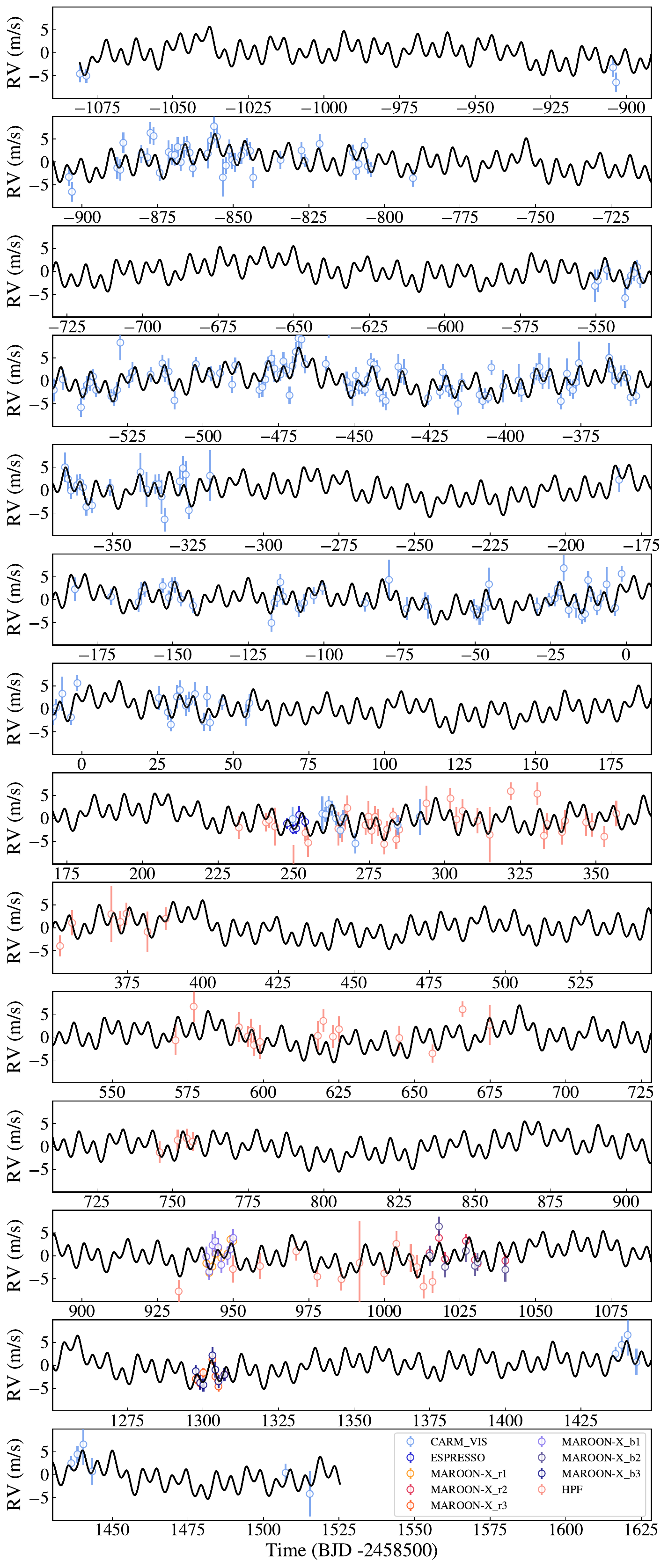}
    \caption{Full dataset overlayed with the best-fit model (model E, see Table\,\ref{tab:modselect}). The error bars contain the jitter contribution listed in Table\,\ref{tab:modresult_d}.}
    \label{fig:RV_time}
\end{figure}

\begin{figure}
    \centering
    \includegraphics[width=0.49\textwidth,page=2]{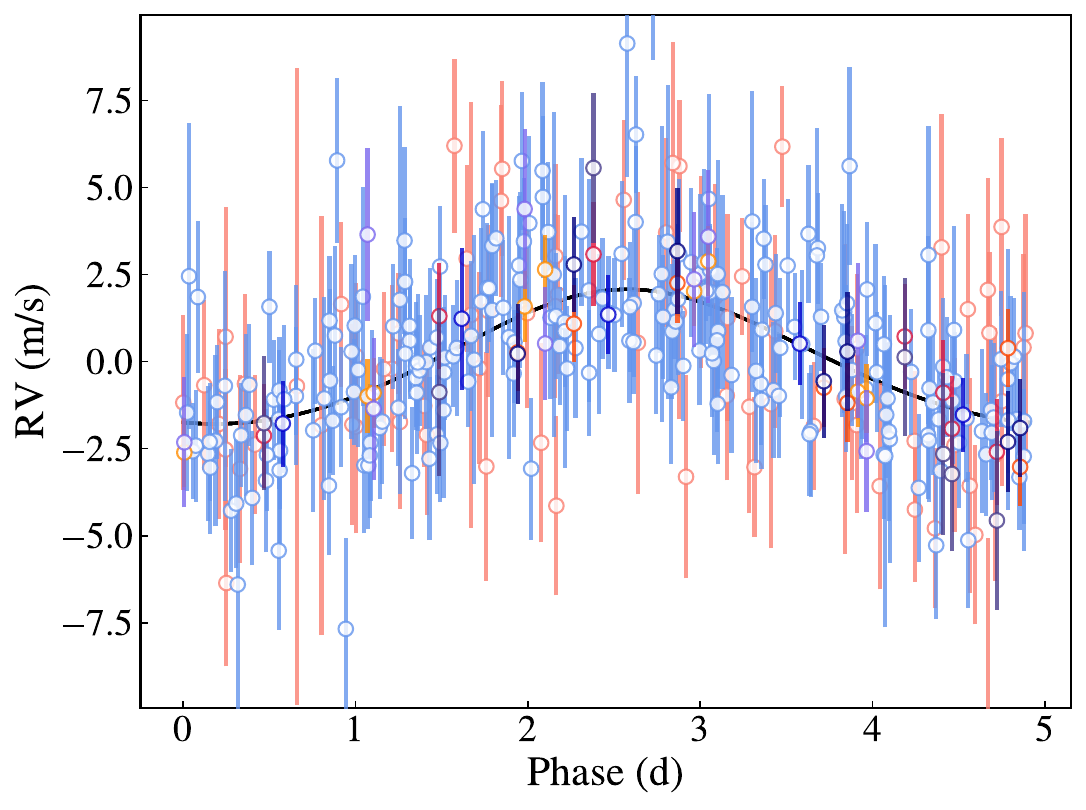}
    \includegraphics[width=0.49\textwidth,page=3]{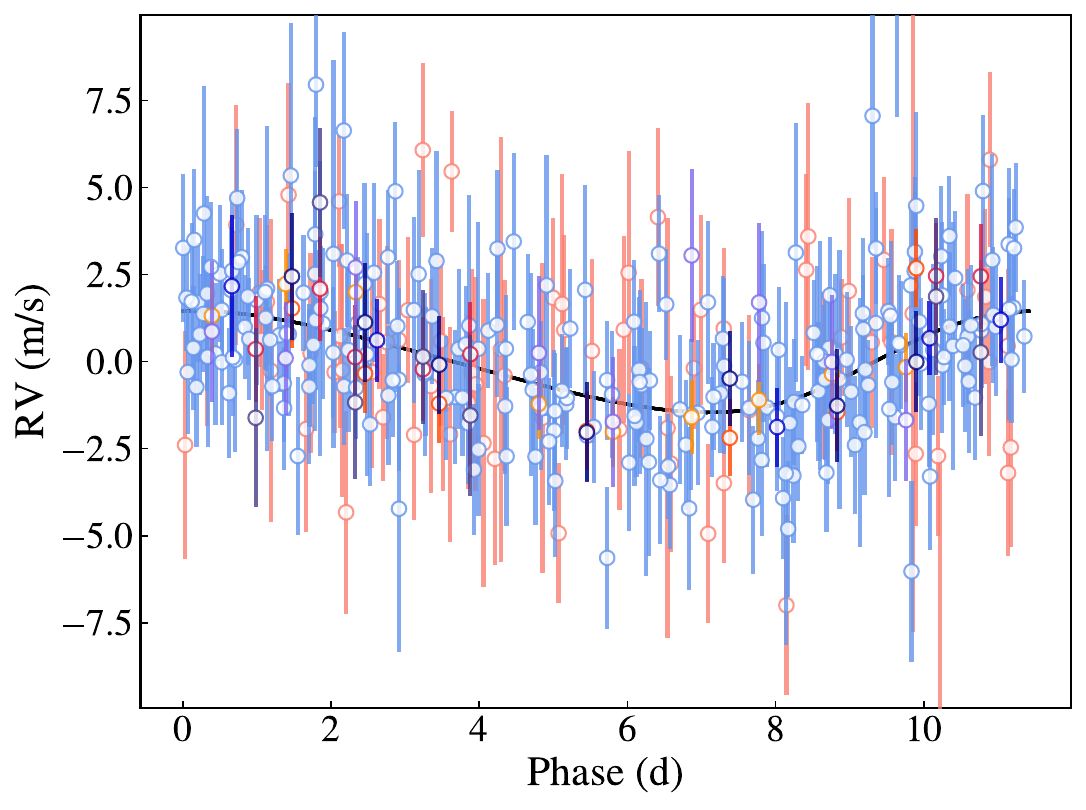}
    \includegraphics[width=0.49\textwidth,page=4]{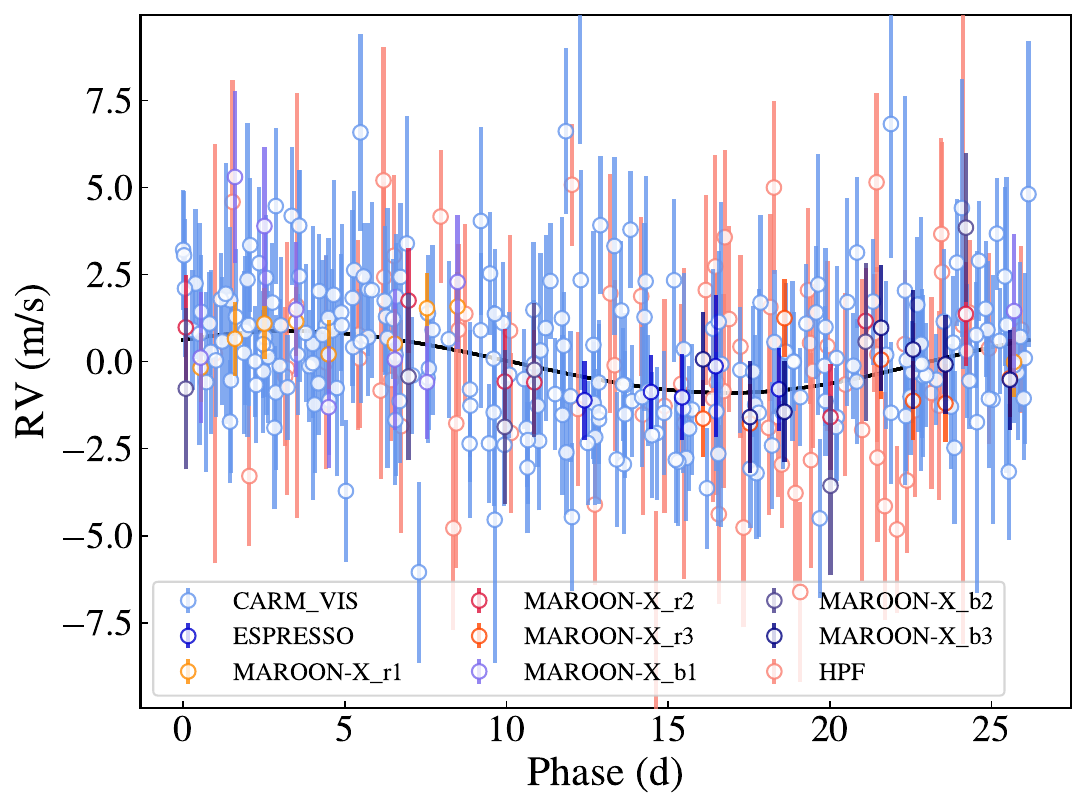}
    \caption{Phase-folded full dataset overlayed with the best-fit model (model E, see Table\,\ref{tab:modselect}). From top to bottom: Planet b, c, d. For each panel, the contribution from the other signals has been filtered out. The error bars contain the jitter contribution listed in Table\,\ref{tab:modresult_d}.}
    \label{fig:RV_phase}
\end{figure}

The adopted fits are presented in Figs.\,\ref{fig:RV_time} and \ref{fig:RV_phase} over time in intervals of half a year and in orbital phase for each planetary signal, respectively. The parameters and their uncertainties as well as the priors are listed in Tables\,\ref{tab:modresult_p} and \ref{tab:modresult_d} for the planetary and data parameters, respectively. 
Table\,\ref{tab:modresult_p} also contains a selection of derived planetary parameters, such as the minimum planetary mass, the equilibrium temperature (assuming an albedo of 0.3), and the instellation. In the Appendix, we also list the parameters for the additional two signals that are likely due to stellar activity but better represented by Keplerian functions (Table\,\ref{tab:modresult_c}). For the 172\,d signal we also list a minimum planetary mass in case the further observations reveal a planetary origin of this signal (and it would become Teegarden's Star e). Finally, the posterior distribution and correlations for all fit parameters are displayed in Figs.\,\ref{fig:corner} to \ref{fig:corner_d}. 

\begin{table}
    \centering
    \caption{Fit and derived planetary parameters (model E).}
    \begin{tabular}{@{}lr@{}ll@{}}
        \hline
        \hline
        \noalign{\smallskip}
        Parameter  & \multicolumn{2}{c}{Posterior}& Prior distribution\\
        \noalign{\smallskip}
         \hline
         \multicolumn{4}{c}{Planet b}\\
         $P$ [d]                & 4.90634& $^{+0.00041}_{-0.00041}$& $\mathcal{U}\, [4.900,4.915]$\\
         $K$ [m\,s$^{-1}$]              & 2.09  &$^{+0.15}_{-0.15}$     & $\mathcal{U}\, [0,3]$\\
         $h$                    &$-$0.03  & $^{+0.15}_{-0.15}$    & $\mathcal{N}\, [0,0.45, -1, 1]$\\
         $k$                    & +0.01  & $^{+0.15}_{-0.14}$    & $\mathcal{N}\, [0,0.45, -1, 1]$\\
         $\lambda$ [deg]         & 171.8 & $^{+4.0}_{-4.0}$      & $\mathcal{U}\, [0,360]$\\
         \noalign{\medskip}
         $m\,\sin i$ [M$_\oplus$]& 1.16  & $^{+0.12}_{-0.11}$ &  \\
         $a$ [au]               & 0.0259&$^{+0.0008}_{-0.0009}$ &  \\
         $e$                    & 0.03 &$^{+0.04}_{-0.02}$      &  \\
         $\omega$ [deg]         & 338 &$^{+133}_{-100}$          &  \\
         $T_\mathrm{eq}$ [K] & 277 & $^{+5}_{-5}$ & \\
         $S$ [S$_\odot$] & 1.08& $^{+0.08}_{-0.08}$ & \\
         \noalign{\smallskip}
         \hline
         \multicolumn{4}{c}{Planet c}\\
         $P$ [d]                & 11.416& $^{+0.003}_{-0.003}$  & $\mathcal{U}\, [11.37,11.44]$\\
         $K$ [m\,s$^{-1}$]              & 1.43  &$^{+0.16}_{-0.15}$     & $\mathcal{U}\, [0,3]$\\
         $h$                    &$-$0.07  & $^{+0.17}_{-0.19}$    & $\mathcal{N}\, [0,0.45, -1, 1]$\\
         $k$                    &$-$0.02  & $^{+0.16}_{-0.16}$    & $\mathcal{N}\, [0,0.45, -1, 1]$\\
         $\lambda$ [deg]        & 336.7 & $^{+6.2}_{-6.3}$      & $\mathcal{U}\, [0,360]$\\
         \noalign{\medskip}
         $m\,\sin i$ [M$_\oplus$]& 1.05& $^{+0.14}_{-0.13}$  &  \\
         $a$ [au]               & 0.0455&$^{+0.0015}_{-0.0016}$&  \\
         $e$                    & 0.04 &$^{+0.07}_{-0.03}$     &  \\
         $\omega$ [deg]         & 301   &$^{+165}_{-74}$       &  \\
         $T_\mathrm{eq}$ [K] & 209 & $^{+4}_{-4}$ & \\
         $S$ [S$_\odot$] & 0.35& $^{+0.02}_{-0.02}$ & \\
         \noalign{\smallskip}
         \hline
         \multicolumn{4}{c}{Planet d}\\
         $P$ [d]                & 26.13 & $^{+0.03}_{-0.04}$    & $\mathcal{U}\, [25.70,26.30]$\\
         $K$ [m\,s$^{-1}$]              & 0.86  &$^{+0.17}_{-0.17}$     & $\mathcal{U}\, [0,2]$\\
         $h$                    &$-$0.05  & $^{+0.20}_{-0.21}$    & $\mathcal{N}\, [0,0.45, -1, 1]$\\
         $k$                    & +0.04  & $^{+0.23}_{-0.22}$    & $\mathcal{N}\, [0,0.45, -1, 1]$\\
         $\lambda$ [deg]         & 325.7 & $^{+11.8}_{-11.3}$    & $\mathcal{U}\, [0,360]$\\
         \noalign{\medskip}
         $m\,\sin i$ [M$_\oplus$]& 0.82& $^{+0.17}_{-0.17}$     &  \\
         $a$ [au]                & 0.0791&$^{+0.0025}_{-0.0027}$&  \\
         $e$                     & 0.07  &$^{+0.10}_{-0.05}$    &  \\
         $\omega$ [deg]         & 345   &$^{+129}_{-93}$        &  \\
         $T_\mathrm{eq}$ [K] & 159 & $^{+3}_{-3}$ & \\
         $S$ [S$_\odot$] & 0.12& $^{+0.01}_{-0.01}$ & \\
         \noalign{\smallskip}
         \hline
         \noalign{\smallskip}
   \end{tabular}
    \label{tab:modresult_p}
\end{table}

As a final check, we followed the approach of \citet{2023AJ....166...62B} and tested the robustness of the final model against overfitting. 80\% of the RV measurements of each dataset were used to train model E again. The remaining 20\% of the data were kept as control set. Using three random selections of the training set the combined weighted {\it rms} of the residuals of the training and control set, both calculated with the best-fit model parameters derived from the training set, are nearly identical (1.79\,m\,s$^{-1}$ versus 1.78\,m\,s$^{-1}$). This indicates a robust model with reliable predictive properties. This is not unexpected since adding new RV data left all planet parameters unchanged as already shown in Fig.\,\ref{fig:hist_planets}.

\subsection{Transit search and detection limits}
\label{sect:transits}

In this subsection, we aim to confirm or refute the transiting nature of the planets orbiting Teegarden's Star. To this end, we explored the public data provided by the {\em TESS} mission, which observed this star in sectors 43, 70, and 71 during the first mission extension in September 2021 and September and October 2023 using the 2\,min and 10\,min cadences. For our study, we used only the data corresponding to the 2\,min cadence.

Neither the Science Processing Operations Center \citep[SPOC;][]{2016SPIE.9913E..3EJ} nor the Quick-Look Pipeline \citep[QLP; ][]{2020RNAAS...4..204H} have issued an alert for Teegarden's Star, which is a hint of the non-transiting nature of this system. However, these pipelines may not detect some periodic transits if they are shallow and their S/N are below their detection thresholds. Then, the community often conducts complementary planetary searches either using space- and ground-based telescopes (see e.g., \citealt{2022A&A...667A..59D, 2023Natur.617..701P, 2021Sci...371.1038T}) or using alternative custom detection pipelines (see e.g., \citealt{2023MNRAS.518L..31M}). 

In this context, we scrutinized the {\em TESS} data employing the {\tt SHERLOCK} pipeline\footnote{ {\tt SHERLOCK} ({S}earching for {H}ints of {E}xoplanets f{R}om {L}ightcurves {O}f spa{C}e-based see{K}ers) code is fully available on GitHub: \url{https://github.com/franpoz/SHERLOCK}}, presented initially by \cite{2020A&A...641A..23P} and \cite{2020A&A...642A..49D}, and used in several studies (see e.g., \citealt{2021A&A...653A..97W,2021A&A...650A.205V,2022A&A...657A..45S}). This pipeline allows the user to recover known planets, identify planetary candidates, and to search for new periodic signals, which may hint at the existence of additional transiting planets. In short, {\tt SHERLOCK} combines several modules to (i) download and prepare the light curves from the MAST repository, (ii) search for planetary candidates, (iii) perform a semi-automatic vetting, (iv) compute a statistical validation, (v) model the signals to refine their ephemerides, and (vi) compute observational windows for ground-based observatories to trigger a follow-up campaign. We refer the reader to \cite{2023A&A...672A..70P} for a recent application and further details. 

Our first investigation covered potential orbital periods from 0.5\,d to 13\,d. This range guarantees the occurrence of at least two transits for any transiting planet. Simultaneously, it provides ample opportunity to identify planets b and c, in the hypothetical case that they transited. In a second trial, we searched for single transit-like events, which covered the potential scenario where the outermost planet d transits. In neither of these two searches, we found any hint of a signal that might correspond to a transiting planet in the system. All the signals detected were attributable to systematics or noise. 

In view of this negative result, we stressed the {\em TESS} data to test if the photometric precision is sufficient to detect the presence of planets b and c, or any other planets in close-in orbits with masses below the detectability limit of our RVs.
For this purpose, we used the {\tt MATRIX} \footnote{{The {\tt MATRIX (\textbf{M}ulti-ph\textbf{A}se \textbf{T}ransits \textbf{R}ecovery from \textbf{I}njected e\textbf{X}oplanets}) code is open access on GitHub: \url{https://github.com/PlanetHunters/tkmatrix}}} code \citep{2022zndo...6570831D, 2020A&A...641A..23P} to search for small, nearby planets using an injection and retrieval test. The simulation was set up with 24 period bins between 0.5\,d and 12\,d, as well as 15 planetary radius bins between 0.5\,R$_\oplus$ and 1.3\,R$_\oplus$. Each synthetic planet in this period-radius parameter space was injected at four different phases. The data were detrended with Tukey's biweight (or bisquare) method \citep{mosteller1977data} with a 0.5\,d window size. As a result, planets larger than 0.5\,R$_\oplus$ would be detected with a detection efficiency close to 100\% in orbits shorter than 7\,d (yellow area in Fig.\,\ref{fig:recovery}) and with an efficiency better than about 75\% in the period range of 7\,d to 12\,d. Using the consecutive sectors 70 and 71 we conducted a similar test for 15 periods between 25\,d and 27\,d, 15 radius bins between 0.5\,R$_\oplus$ and 1.3\,R$_\oplus$ and 4 phase bins (Fig.\,\ref{fig:recovery_26}. Planet d would have been detected with high probablity if it would be a transiting planet.

A non-detection in the {\em TESS} data, therefore, excludes transiting planets with most of the parameters that we tested, especially small planets in close-in orbits. We should have found planet b at 4.9\,d and c at 11.4\,d orbital periods if they transited. Planet d with a 26\,d orbital period would at most show one transit in a single {\em TESS} sector and up to two in the two consecutive ones. However, due to the lower geometric transit probability, planet\,d is unlikely to be transiting, while b and c are not. 
Combined together, these results confirm the non-transiting nature of any planet with an orbital period shorter than 12\,d in the Teegarden's Star system, in particular the planets\,b and c, supporting the preliminary results from ground-based data by \citet{2019A&A...627A..49Z}.

\subsection{Dynamical stability}
\label{sect:stability}

We used {\tt SPOCK} to check the resulting planetary system configurations for dynamical stability.
We used the posterior distributions and not draws from the normal distributions given by the mean and standard deviation shown in Table\,\ref{tab:modresult_p}. This analysis reveals that all parameter configurations drawn from the posteriors are dynamically stable. This is not surprising, since the planetary system is not dynamically compact as further discussed below (Sect.\,\ref{sect:discussion}). In case of the confirmation of the 7.7\,d signal (Sect.\,\ref{sect:select}) as the orbital period of an additional planet, the dynamical stability of the planetary system would then require small eccentricities in order to prevent close encounters.
 
\section{Discussion}
\label{sect:discussion}

\subsection{Planet, stellar activity, or instrumental artifact}

As already mentioned in Sect.\,\ref{sect:select}, the signal at 96\,d is likely associated with stellar activity and we interpret it as the potential stellar rotation period \citep{2021A&A...652A..28L}.
Also as described in Sect.\,\ref{sect:results}, the nature of the 172\,d signal is unclear. On the one hand, if the rotation period is 96\,d, 172\,d is sufficiently separated from twice the rotation period, and likewise if 96\,d was assumed to be the 1-year alias of a 79\,d rotation period. Since the signal is coherent over the duration of the observations, the activity-induced variability would need to be very stable in time. Normally, a coherent signal without a harmonic close to the potential rotation period or to its alias would be an indication of a planetary signal. On the other hand, the existence of variability with very similar period in spectroscopic activity indices is a strong indicator of an intrinsic stellar effect rather than a Keplerian orbital motion. The wavelength dependence of the amplitude does not provide any further information to discern the signal's nature. The long-duration datasets that expand the wavelength coverage, CARMENES NIR and HPF, have insufficient precision to determine the amplitude of the 172\,d signal and investigate its potential wavelength dependence. The conservative interpretation for the 172\,d signal therefore is an activity related signal, probably not connected to the stellar rotation period. 

The period of planet d at $26.13\pm 0.04$\,d falls comfortably outside the lunar sidereal and synodic periods of 27.32\,d and 29.53\,d, respectively, which generate small peaks in the periodograms of the spectroscopic activity indicators \citep{kemmer}. To investigate the occurrence of common periodicities detected in all activity indicators of CARMENES stars, Kemmer et al. use a clustering algorithm. A cluster close to the sidereal and synodic month was detected in 5 out of 136 stars. This could indicate that the preferential scheduling of CARMENES observations in bright or gray time imprints the lunar cycle
into the window function of the sampling in some cases. The distinction between the period of planet d and the sidereal month assures that there are no concerns regarding the interpretation of the 26.13\,d signal as planetary in nature. It also does not coincide with a harmonic of the rotation period nor is it affected by spectral leakage of one of the other signals (Fig.\,\ref{fig:simulation}). We therefore conclude that this signal is a bona fide planetary RV signal. 

\subsection{Habitability}

With the minor adjustments in stellar and planetary parameters when compared to \citet{2019A&A...627A..49Z} and with a new planet in the system, it is worthwhile to reevaluate the habitability of these planets. Teegarden's Star planet b continues to exhibit Earth-like characteristics, with an equilibrium temperature of 277\,K, assuming an albedo of 0.3, and an instellation (1.08 S$_\odot$) very close to that of our Earth by the Sun. The Earth Similarity Index \citep[ESI][]{2011AsBio..11.1041S} has only slightly decreased to 0.90, no longer holding the highest ESI ranking according to the habitable exoplanet catalog\footnote{\url{https://phl.upr.edu/projects/habitable-exoplanets-catalog}}. While planet b has experienced a marginal ESI downgrade, planet c now has an ESI of 0.88, closely resembling Proxima\,b \citep{2016Natur.536..437A}. In contrast, the newly discovered planet d is cold, residing on an orbit of about a month, resulting in temperatures akin to Jupiter or its icy moon Ganymede. The prospect of directly detecting the planet's atmosphere, which may be feasible with future instruments given the relatively short distance to Teegarden's Star, becomes profoundly intriguing. Such a discovery would offer the opportunity to investigate the atmospheres of small rocky planets across a wide range of surface temperatures.
\subsection{Planetary system architecture}

\begin{figure*}
    \centering
    \includegraphics[width=0.99\textwidth]{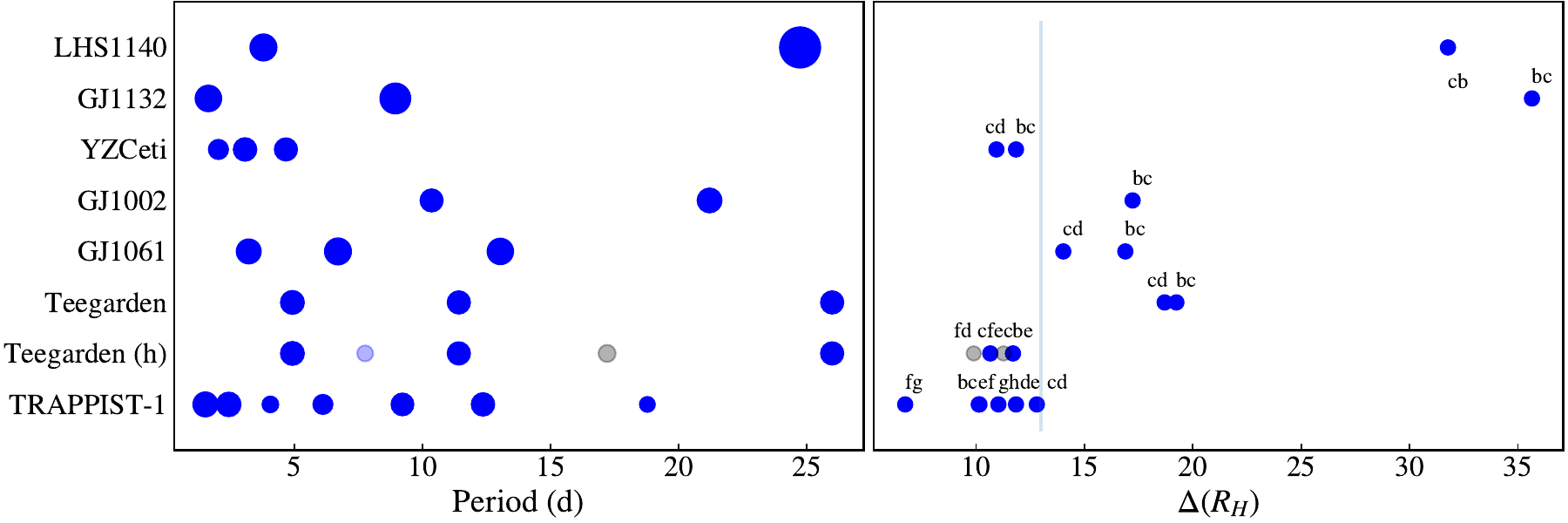}
    \caption{Planetary system architectures in comparison. {\em Left panel}: Semi-major axes in planetary systems in comparison. For Teegarden's Star we show two possible configurations: the three planets as fit by our preferred model and a hypothetical configuration (labeled as Teegarden (h)) with the three planets, the candidate at 7.7\,d (light blue) as well as a purely hypothetical planet between planets c and d (light gray). The size of the plot symbols in the left panel are scaled assuming a constant planet density using the masses from Table\,\ref{tab:modresult_p} and \ref{tab:modresult_c}. {\em Right panel}: Mutual Hill separations in planetary systems in comparison. The light blue line indicates the mutual Hill radius above which the orbital crossing time is larger than 10$^9$ orbits.}
    \label{fig:architecture}
    \label{fig:hill}
\end{figure*}

We now compare the planetary system architecture of Teegarden's Star to that of similar planetary systems. From the NASA Exoplanet archive\footnote{\url{https://exoplanetarchive.ipac.caltech.edu/}} we  extracted all systems fulfilling the following criteria: (i) more than one known planet, (ii) host star with $M_{\rm star} < 0.2 $\,M$_{\odot}$, (iii) planet mass measured from RVs or transit timing variations (i.e., dynamical mass), and (iv) at least one potential rocky planet $M_{\rm planet} < 2 $\,M$_{\oplus}$. This yielded a total of seven systems including Teegarden's Star, namely the following sorted by decreasing host star mass (taken from the references of the default parameter set as listed in the exoplanet archive): 
LHS\,1140 (c[+b]; \citealt{2020A&A...642A.121L}), GJ\,1132 (b[+c]; \citealt{2018A&A...618A.142B}), YZ\,Cet (bcd; \citealt{2020A&A...636A.119S}), GJ\,1002 (bc; \citealt{2023A&A...670A...5S}), GJ\,1061 (bcd; \citealt{2020MNRAS.493..536D}), and TRAPPIST-1 (bcdefgh; \citealt{2021PSJ.....2....1A}). 

The planetary system architectures are displayed in Fig.\,\ref{fig:architecture}, where the size of the plot symbols scales with the planet radius derived assuming Earth-like bulk density. We note that the minimum planet masses are used except for TRAPPIST-1. 
These systems are all compact and their planets orbit within 0.1\,au. Only the two with the highest host star mass, LHS\,1140 and GJ\,1132, have planets above $M_{\rm planet} > 2 $\,M$_{\oplus}$. A good indicator of the dynamic compactness of a planetary system is the mutual Hill radius also termed fractional orbital separation \citep{1993Icar..106..247G}. Following \citet{2017Icar..293...52O}, it is calculated from the stellar mass $M_{\rm star} \equiv M$, the planetary masses $M_{\rm planet} \equiv m$, and semi-major axes $a$ of two neighboring planets $j$ and $j+1$ as

\begin{equation*}
    \Delta = 2\frac{a_{j+1}-a_j}{a_{j+1}+a_j}\left(\frac{3M}{m_{j+1}+m_j}\right)^{1/3}.
\end{equation*}
Using the mutual Hill radii (Fig.\,\ref{fig:hill}) also shows that the LHS\,1140 and GJ\,1132 systems seem different compared to the ones with lower host star mass. The planets of systems with more massive planets have mutual Hill radius separations above $\Delta \gtrsim 30$ while the ones with low-mass, likely terrestrial, planets have $\Delta \lesssim 20$. 

The mutual Hill separations can be compared to those derived from multi-planet systems detected by {\em Kepler} \citep[e.g.,\,][their Fig. 14]{2018AJ....155...48W}. The median mutual Hill separation of two-planet {\em Kepler} systems, whose hosts are predominantly FGK stars, is about 25. The planets of LHS\,1140 and GJ\,1132 fall within this distribution. The results by \citet{2018AJ....155...48W} also indicate that the mutual Hill radii tend to be smaller if more planets are present, but only about 10\,\% of the systems have a $\Delta < 10$. While the stability of two-planet systems can be described analytically \citep{0eb94dbb-e87d-35f4-a82b-f98b26a6e529,1993Icar..106..247G}, the investigation of the orbital stability of systems with more planets requires extensive numerical simulations \citep[e.g.,][]{1996Icar..119..261C,2009Icar..201..381S,2017A&A...604A...1O,2021Icar..35814038G,2023MNRAS.520.4057R}. 
A common result is that the logarithm of the time of the first orbital crossing depends linearly on the mutual Hill radius. At about 13 mutual Hill radii, the orbital crossing time scale seems to be longer than at least 10$^9$ orbits for all systems. Below this limit the orbital crossing time scale changes by more than one order of magnitude for small changes in mutual Hill separation. These fluctuations are due to first and second order period commensurabilities, with additional dependencies on eccentricities and planet mass ratios.

The mutual Hill radii (Fig.\,\ref{fig:hill}), however, reveal that there seem to be two types of configurations. All systems except TRAPPIST-1 and YZ\,Ceti have mutual Hill radius separation above the threshold $\Delta \gtrsim 13$ derived from numerical simulations where the orbit crossing time scale is probably at least as long as the system age. The planetary system of Teegarden's Star, as obtained from our analysis with three planets, is well above this threshold for the two adjacent planet pairs (Teegarden label in Fig.\,\ref{fig:hill}). A putative planet at about 7.7\,d would, however, change this picture drastically. Planets b and c would then form pairs with this potential planet e with mutual Hill separations only slightly above 10, very similar to planet pairs in the TRAPPIST-1 system (Teegarden (h) in Fig.\,\ref{fig:hill}). The signal that we investigated in Sect.\,\ref{sect:CARMENES} at 7.7\,d would correspond to a planet below 0.5\,M$_\oplus$ and an RV amplitude of about 0.5\,m\,s$^{-1}$. Given the measurement uncertainties, such a low-amplitude signal is at or below our detection limit but not at all unrealistic given the masses of the TRAPPIST-1 planets, which range from 0.33\,M$_\oplus$ to 1.37\,M$_\oplus$ \citep{2021PSJ.....2....1A}. A hypothetical low-mass planet between planets c and d at about 17\,d would be even more difficult to detect, but would turn the planetary system of Teegarden's Star into a closer, and therefore brighter, twin of the TRAPPIST-1 system, with similar very low stellar mass, and multiple low-mass planets in a tightly packed system. More stringent upper limits on missing planets  could, on the other hand, help to better establish the alternative, namely a rather loosely packed architecture of the planetary system of Teegarden's Star. 

\subsection{Stellar flares observed using SPECULOOS telescopes}
\label{sect:speculoos}
A detailed inspection of the SPECULOOS data confirmed the non-transiting nature of the system, supporting the results obtained in Section\,\ref{sect:transits} using {\em TESS} data. However, these observations revealed several stellar flares. We identified 13 stellar flares by eye: 10 clear, single-peak flares and one multi-flare region with 3 peaks. There was an additional large flare with a slow, extended tail that could not be satisfactorily fit with the \citet{2016ApJ...829...23D} flare model or combination of flares and was not considered in this flare study. To obtain accurate flare energies we used \textsc{PyMC3}'s MCMC sampler \citep{2015arXiv150708050S} and xoflares \citep{Gilbert2022, 2016ApJ...829...23D} to model the flares simultaneously with linear systematic models for the full width half-maximum (FWHM), change in x and y CCD positions, sky background and airmass. We followed the method outlined in \citet{2022MNRAS.513.2615M} to calculate the flaring rates and U-band energies and generate the flare frequency diagram (FFD) for Teegarden's Star. We do not account for the sensitivity of our flare detection as we are considering a single object. The reduced detection sensitivity at low flare energies will result in a tail-off (or flattening of the FFD) at low energies, however, it is not visible in the FFD in this case. The power law $dN(E) = kE^{-\alpha}dE$ \citep{Gershberg1972, Lacy1976, 2014ApJ...797..121H} was fit to the FFD using \textsc{PyMC3}, where $N$ is the flare occurrence rate, $E$ is the flare energy, and $k$ and $\alpha$ are constants. We obtained a best-fit $\alpha$ of 1.84$\pm$0.05, consistent with several sample studies of mid-to-late M dwarfs \citep{2017ApJ...838...22G, Paudel2018, Raetz2020, 2022MNRAS.513.2615M}. The FFD indicates that Teegarden's Star may produce flares energetic enough for a planet with 1.08 Earth irradiation, such as planet b (Table \ref{tab:modresult_p}), to enter the abiogenesis zone defined by \citet{2018SciA....4.3302R}. On average, flare energies $\gtrsim 10^{35}$ergs will provide enough UV energy to build up a prebiotic inventory for RNA synthesis on planet b and, by extrapolating from SPECULOOS's low energy flare sample, these type of flares should occur at most once every 2.4 years.

\section{Summary}
\label{sect:summary}

Through a reanalysis of 346 nightly binned RV data collected from CARMENES, ESPRESSO, MAROON-X, and HPF, we report the discovery of a third planet in the Teegarden's Star system. This newly identified planet, Teegarden's Star d, has an orbital period of $26.13\pm0.04$\,d. The RV amplitude of $0.86\pm 0.17$\,m\,s$^{-1}$ corresponds to a minimum mass of 0.82\,M$_\oplus$. However, due to the very low mass of the primary star, this planet orbits outside the habitable zone of Teegarden's Star. 

We have also observed an additional signal at 96\,d, which is in good agreement with the stellar rotation period previously reported by \citet{2021A&A...652A..28L}. The nature of another signal at 172\,d, with an amplitude of 1.3\,m\,s$^{-1}$, remains unclear. It can be attributed to stellar activity on a timescale longer than the rotation period or potentially indicate the presence of a long-period planet with a minimum mass of 2.3\,M$_\oplus$. However, the coincidence of the period with spectroscopic indicators makes the planet hypothesis less likely. Moreover, we found evidence of very long-period variability, spanning around 2\,500\,d or longer, in the RV data and spectroscopic activity indicators, which may indicate the presence of an activity cycle.

In the CARMENES data, two additional signals of modest significance at a level of 0.5\,m\,s$^{-1}$ were also identified. However, they are not detectable in the full dataset. One of these signals, with a period of 7.7\,d, would correspond to a planet with approximately half the mass of Earth, situated close to a 3:2 commensurability chain with planets b and c. If confirmed, this planet would make the Teegarden's Star planetary system tightly packed, resembling the TRAPPIST-1 system. A thorough search for yet undetected sub-Earth mass planets is crucial to elucidate the architectural characteristics of nearby planetary systems, particularly those suitable for atmospheric characterization.

\begin{acknowledgements}
We thank the referee for constructive and timely comments.

We acknowledge the support from the Deutsche Forschungsgemeinschaft (DFG) under Research Unit FOR\,2544 "Blue Planets around Red Stars" through project DR~281/32-1.
 
  This publication was based on observations collected under the CARMENES Legacy+ project.
  CARMENES is an instrument at the Centro Astron\'omico Hispano en Andaluc\'ia (CAHA) at Calar Alto (Almer\'{\i}a, Spain), operated jointly by the Junta de Andaluc\'ia and the Instituto de Astrof\'isica de Andaluc\'ia (CSIC).
  CARMENES was funded by the Max-Planck-Gesellschaft (MPG), 
  the Consejo Superior de Investigaciones Cient\'{\i}ficas (CSIC),
  the Ministerio de Econom\'ia y Competitividad (MINECO) and the European Regional Development Fund (ERDF) through projects FICTS-2011-02, ICTS-2017-07-CAHA-4, and CAHA16-CE-3978, 
  and the members of the CARMENES Consortium 
  (Max-Planck-Institut f\"ur Astronomie,
  Instituto de Astrof\'{\i}sica de Andaluc\'{\i}a,
  Landessternwarte K\"onigstuhl,
  Institut de Ci\`encies de l'Espai,
  Institut f\"ur Astrophysik G\"ottingen,
  Universidad Complutense de Madrid,
  Th\"uringer Landessternwarte Tautenburg,
  Instituto de Astrof\'{\i}sica de Canarias,
  Hamburger Sternwarte,
  Centro de Astrobiolog\'{\i}a and
  Centro Astron\'omico Hispano-Alem\'an), 
  with additional contributions by the MINECO, 
  the DFG through the Major Research Instrumentation Programme and Research Unit FOR2544 ``Blue Planets around Red Stars'', 
  the Klaus Tschira Stiftung, 
  the states of Baden-W\"urttemberg and Niedersachsen, 
  and by the Junta de Andaluc\'{\i}a.
  
  Based on observations obtained at the international Gemini Observatory, a program of NSF’s NOIRLab, which is managed by the Association of Universities for Research in Astronomy (AURA) under a cooperative agreement with the National Science Foundation on behalf of the Gemini Observatory partnership: the National Science Foundation (United States), National Research Council (Canada), Agencia Nacional de Investigaci\'{o}n y Desarrollo (Chile), Ministerio de Ciencia, Tecnolog\'{i}a e Innovaci\'{o}n (Argentina), Minist\`{e}rio da Ci\^{e}ncia, Tecnologia, Inova\c{c}\~{o}es e Comunica\c{c}\~{o}es (Brazil), and Korea Astronomy and Space Science Institute (Republic of Korea). This work was enabled by observations made from the Gemini North telescope, located within the Maunakea Science Reserve and adjacent to the summit of Mauna Kea. We are grateful for the privilege of observing the Universe from a place that is unique in both its astronomical quality and its cultural significance.

  Based on observations obtained with the Hobby-Eberly Telescope (HET), which is a joint project of the University of Texas at Austin, the Pennsylvania State University, Ludwig-Maximillians-Universitaet M\"unchen, and Georg-August Universitaet G\"ottingen. The HET is named in honor of its principal benefactors, William P. Hobby and Robert E. Eberly.

  We acknowledge the use of public TESS data from pipelines at the TESS Science Office and at the TESS Science Processing Operations Center. Resources supporting this work were provided by the NASA High-End Computing (HEC) Program through the NASA Advanced Supercomputing (NAS) Division at Ames Research Center for the production of the SPOC data products.

  We acknowledge the support of the DFG priority program SPP 1992 “Exploring the Diversity of Extrasolar Planets" (DR 281/37-1, HE 1935/28-1, QU 113/8-1, RE 1664/20-1).

  We acknowledge financial support from the Agencia Estatal de Investigaci\'on (AEI/10.13039/501100011033) of the Ministerio de Ciencia e Innovaci\'on and the ERDF ``A way of making Europe'' through projects 
  PID2021-125627OB-C31,		
  PID2019-109522GB-C5[1:4],	
  PID2021-125627OB-C32,        
  PID2022-137241NB-C43         
and the Centre of Excellence ``Severo Ochoa'' and ``Mar\'ia de Maeztu'' awards to the Instituto de Astrof\'isica de Canarias (CEX2019-000920-S), Instituto de Astrof\'isica de Andaluc\'ia (CEX2021-001131-S) and Institut de Ci\`encies de l'Espai (CEX2020-001058-M).
  
  This work was also funded by the Generalitat de Catalunya/CERCA programme; 
  the Universidad La Laguna and the EU Next Generation through the Margarita Salas Fellowship from the Spanish Ministerio de Universidades under grant UNI/551/2021-May-26;
  {ERASMUS+}, the European Union programme for education, training, youth and sport; 
  the David and Lucile Packard Foundation, the Heising-Simons Foundation, and the Gordon and Betty Moore Foundation; 
  the Gemini Observatory; 
  the National Science Foundation under grant 2108465 and graduate research fellowship DGE~1746045;
  and NASA under grant 80NSSC22K0117 *and* the NASA Hubble Fellowship grant HST-HF2-51519.001-A awarded by the Space Telescope Science Institute, which is operated by the Association of Universities for Research in Astronomy, Inc., for NASA, under contract NAS5-26555.
  The ULiege's contribution to SPECULOOS has received funding from the European Research Council under the European Union's Seventh Framework Programme (FP/2007-2013) (grant Agreement n$^\circ$ 336480/SPECULOOS), from the Balzan Prize and Francqui Foundations, from the Belgian Scientific Research Foundation (F.R.S.-FNRS; grant n$^\circ$ T.0109.20), from the University of Liege, and from the ARC grant for Concerted Research Actions financed by the Wallonia-Brussels Federation. MG is F.R.S-FNRS Research Director. 
  The SPECULOOS-North team gratefully acknowledges financial support from the Heising-Simons Foundation, Dr. and Mrs. Colin Masson and Dr. Peter A. Gilman for Artemis, the first telescope of the SPECULOOS network situated in Tenerife, Spain.
  This work is supported by the Swiss National Science Foundation (PP00P2-163967, PP00P2-190080 and the National Centre for Competence in Research PlanetS).

  This research has made use of the NASA Exoplanet Archive, which is operated by the California Institute of Technology, under contract with the National Aeronautics and Space Administration under the Exoplanet Exploration Program.

\end{acknowledgements}

\bibliographystyle{aa}
\bibliography{Teegarden}

\begin{appendix}

\section{Radial velocity data}

Table\,\ref{tab:RVs} lists the original RV data from CARMENES, ESPRESSO, MAROON-X, and HPF. We note that the CARMENES data slightly differ from \citet{2023A&A...670A.139R}, since we here use spectra cleaned from telluric contamination.

\begin{table}
    \centering
    \caption{Full RV dataset prior to binning and clipping. The full table is available in electronic form at the CDS.}
    \begin{tabular}{@{}ccc@{}c@{}}
        \hline
        \hline
        BJD (d) & RV (m\,s$^{-1}$) & $\sigma_{\rm RV}$\,(m\,s$^{-1}$) & Instrument \\
        \hline
       2457419.283539 &-3.284 &1.624& CARMENES VIS \\
       2457421.417489 &-3.731 &1.228& CARMENES VIS \\
       2457595.635969 &-2.569 &1.742& CARMENES VIS \\
       2457596.649266 &-5.808 &1.890& CARMENES VIS \\
       2457611.634408 &-0.660 &2.142& CARMENES VIS \\
         $\vdots$ & $\vdots$& $\vdots$& $\vdots$ \\
        \hline
    \end{tabular}
    \label{tab:RVs}
    \tablefoot{CDS via anonymous ftp to \url{cdsarc.u-strasbg.fr} (130.79.128.5) or via \url{http://cdsweb.u-strasbg.fr/cgi-bin/qcat?J/A+A/}.}
\end{table}

\section{Results of the $\ell_1$-periodogram analysis}
Fig.\,ref{fig:l1periodogram} shows the $\ell_1$-periodogram of the CARMENES VIS RV data using the best noise representation after cross validation of various covariance matrices. The cross validation results in no additional noise terms.

\begin{figure}
    \centering
    \includegraphics[width=0.49\textwidth]{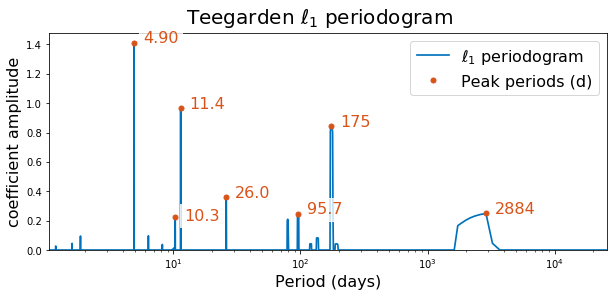}
    \includegraphics[width=0.49\textwidth]{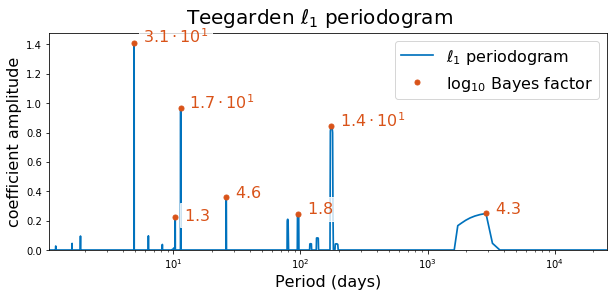}
    \caption{$\ell_1$-periodogram of the CARMENES VIS RV data. The detected signals are identified with the corresponding period in days (top) and with the logarithm of the Bayes factor (bottom).}
    \label{fig:l1periodogram}
\end{figure}

\section{Crosstalk from signals and spectral leakage}

\begin{figure}
    \centering
    \includegraphics[width=0.49\textwidth]{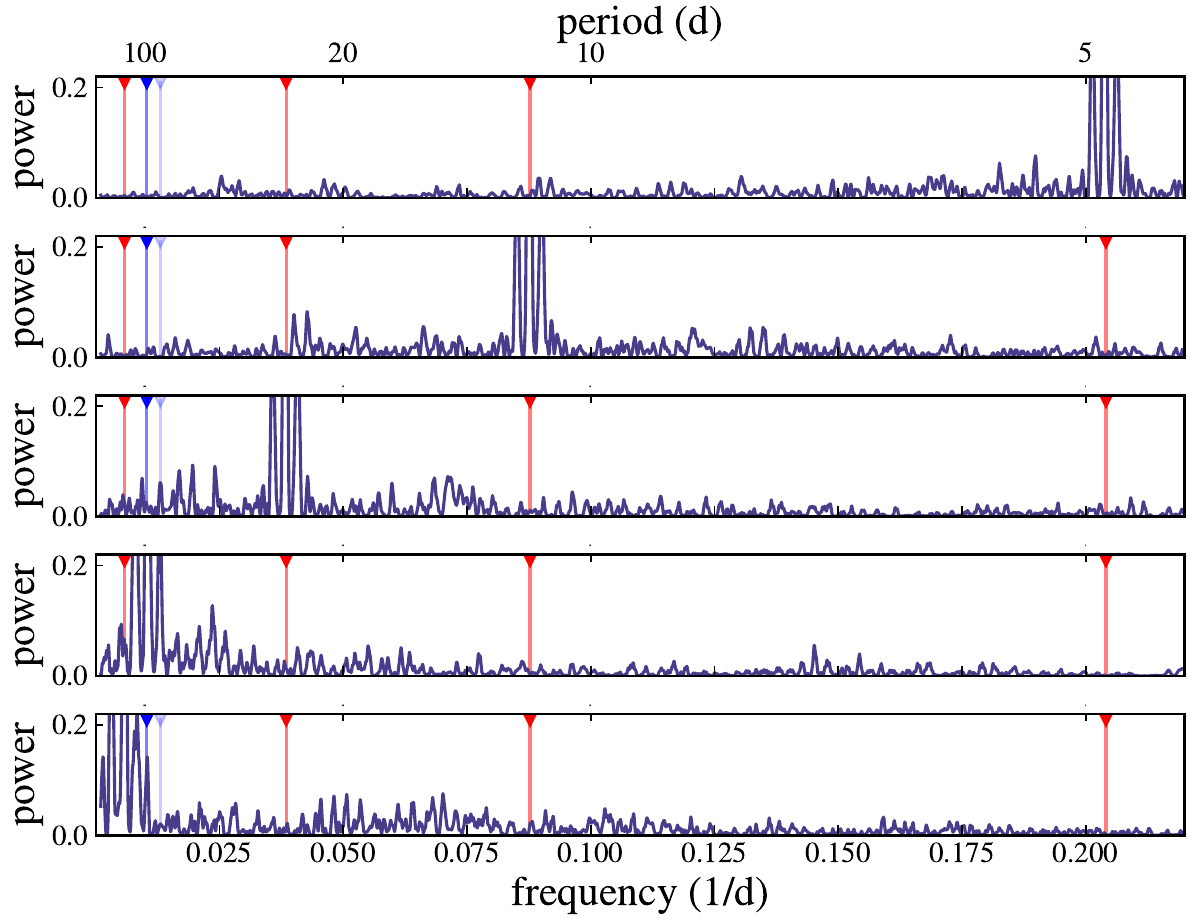}
    \includegraphics[width=0.49\textwidth]{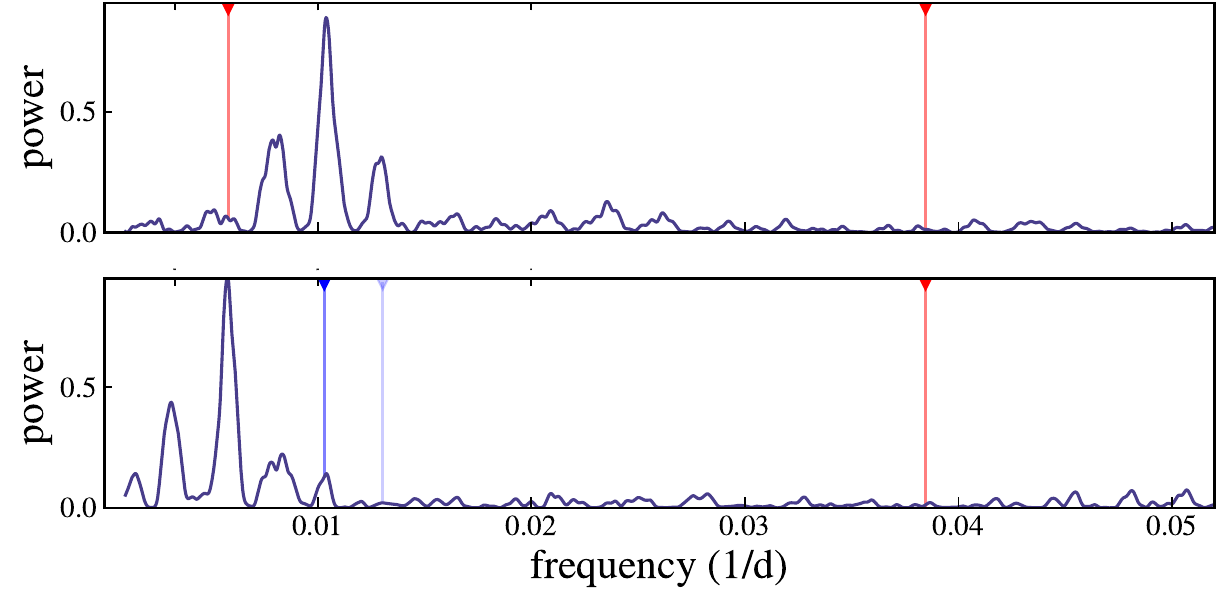}
    \caption{Periodogram of the Keplerian models with the parameters from Table\,\ref{tab:modresult_p}. From top to bottom upper panel: 4.9\,d, 11.4\,d, 26\,d, 96\,d, and 172\,d. The blue triangle indicates the rotation period at 96\,d and its one-year alias at 79\,d, the red triangles indicate the periods of the other signals (4.9\,d, 11.4\,d, 26\,d, and 172\,d). Lower panel: zoom into the long period range for the 96\,d, and 172\,d signals.}
    \label{fig:simulation}
\end{figure}

Due to the complex window function of the time series of the CARMENES observations, we calculated the periodograms of the five signals with the parameters from Table\,\ref{tab:modresult_p} and \ref{tab:modresult_c} with the sampling of the CARMENES observations. The result is depicted in Fig.\,\ref{fig:simulation}.

\section{Signal at 172~d in RV data and spectrospcopic indices}

\begin{figure*}[t]
    \centering
    \includegraphics[width=0.99\textwidth]{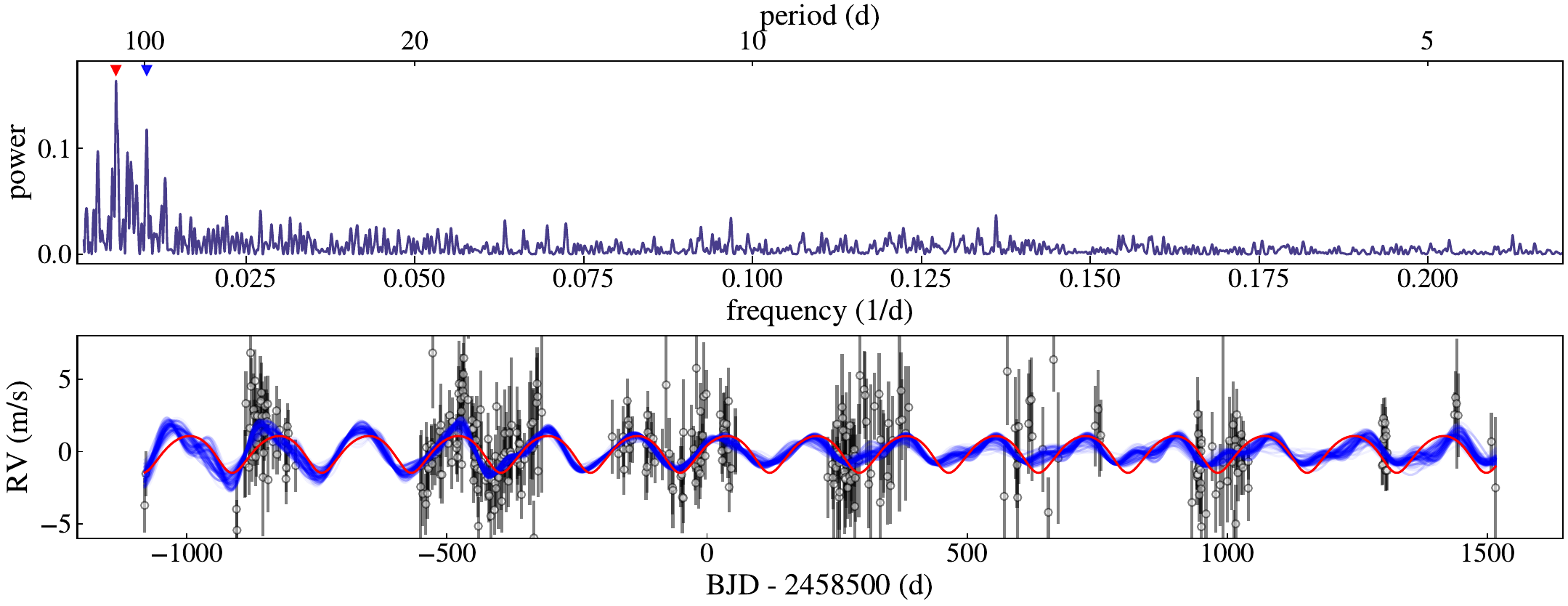}
    \caption{The noise model in frequency and time domain. Top: Periodogram of the correlated noise model of the best-fit of model I. The periodogram shows peaks both at 172\,d (red) and the 96\,d (blue) signal. Bottom: Comparison between the correlated noise model using draws from the posteriors of the hyperparameters for the dSHO kernel from model I (blue) and the best-fit Keplerian model for the 172\,d signal from model E (red). The data points are the offset- and jitter-corrected measurements for all instruments where the contribution from planets b, c, and d have been subtracted.}
    \label{fig:GP_plot}
\end{figure*}

We show the comparison between the correlated noise model using draws from the posteriors of the hyperparameters for the dSHO kernel from model I in blue (Fig.\,\ref{fig:GP_plot} abd the best-fit Keplerian model for the 172\,d signal from model E (red). The data points are the offset-corrected data for all instruments, the error bars are adjusted to the fit jitter value. The contribution of planets b, c, and d have been subtracted. The correlated noise model shows a very long coherence and closely resembles the Keplerian model in amplitude and phase. This similarity of the signals can also be see comparing the $\sqrt{\sigma}_{\rm dSHO}$ hyperparameter in Table\,\ref{tab:modresult_d} with the RV amplitude of the 172\,d signal in Table\,\ref{tab:modresult_c}. The quality factor $Q$ given in Table\,\ref{tab:modresult_d} indicate a damping time scale on the order of twice the observing base line. If interpreted as spot life time, Teegarden's Star would have a spot pattern stable for at least about 4000\,d or 40 rotation periods.

\begin{figure}[h]
    \centering
    \includegraphics[width=0.49\textwidth]{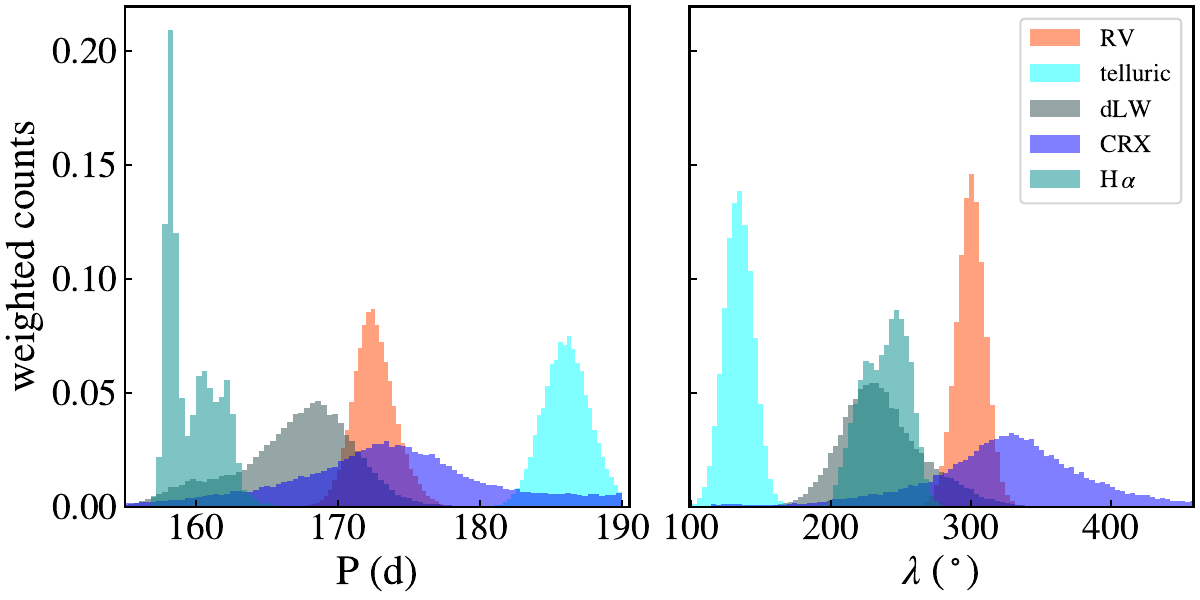}
    \caption{Posterior distribution for a fit of a Keplerian model of the CARMENES RV data (RV) as well as spectroscopic indices: Telluric line contamination (telluric), differential line width (dLW), chromatic index (CRX), and H$\alpha$ index (H$\alpha$). Posteriors for the orbital periods (left) and mean longitudes (right) are shown.}
    \label{fig:hist_indices}
\end{figure}

In order to investigate whether the signal at 172\,d in the RV data is due to stellar activity, instrumental effects, or a planet, we fit various datasets with identical priors and compared the output. In Fig.\,\ref{fig:hist_indices} we show the posterior distributions for the periods and mean longitudes, which can be interpreted as phases in this context.


\section{Stacked Bayesian generalized Lomb-Scargle periodogram}

\begin{figure}
    \centering
    \includegraphics[width=0.49\textwidth]{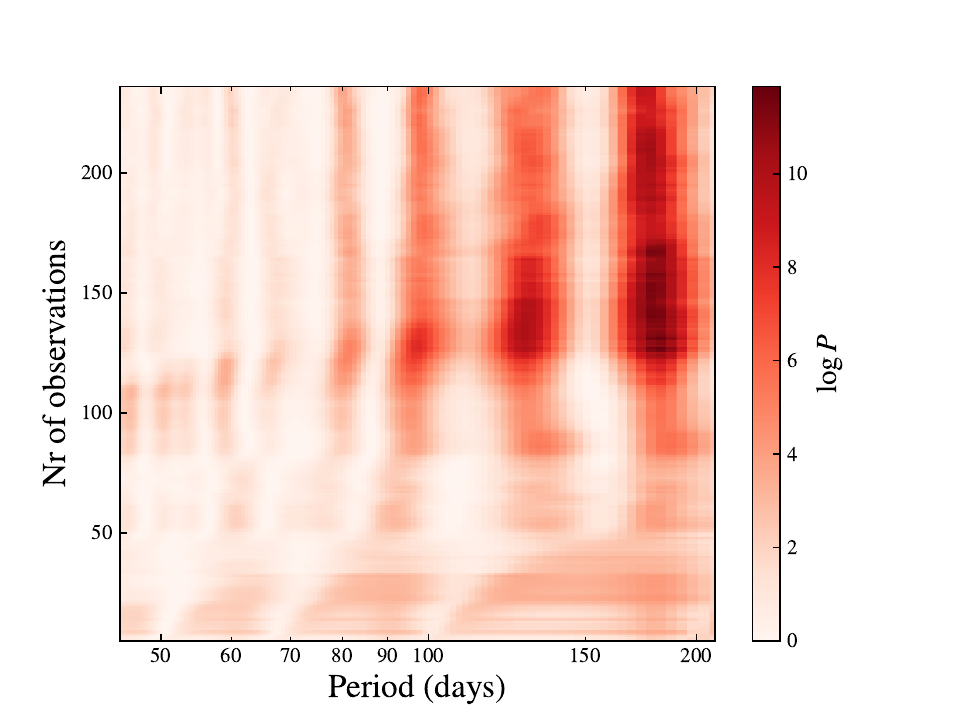}
    \includegraphics[width=0.49\textwidth]{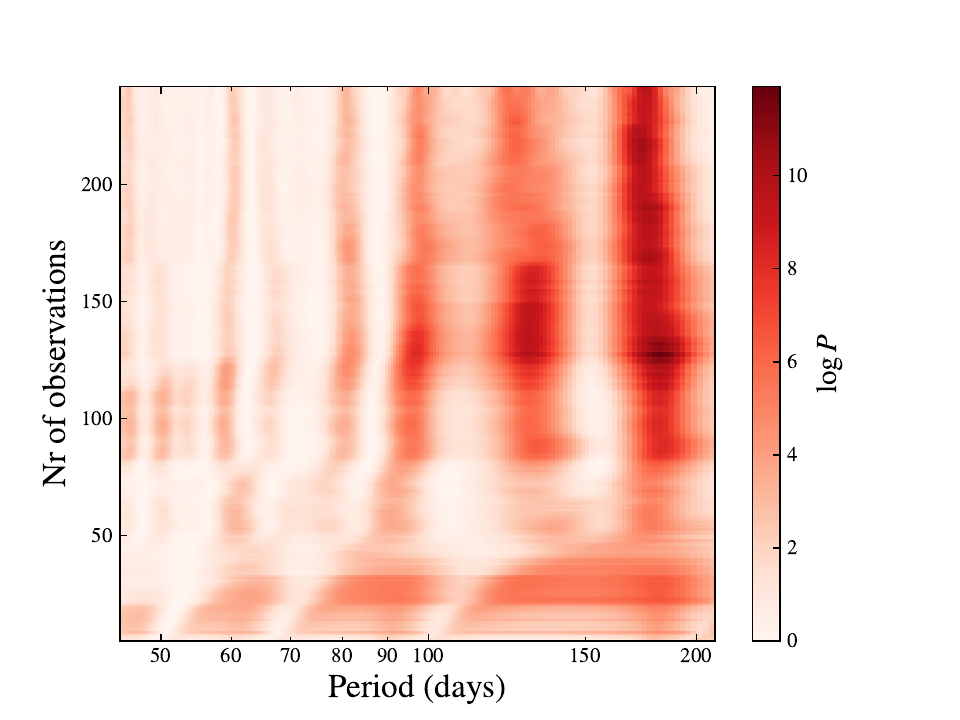}
    \caption{Stacked Bayesian GLS periodogram in the long-period range after subtraction of the signals from Teegarden's Star b and c (model A, top) and of a simulated 172\,d Keplerian orbit (bottom). 
    }
    \label{fig:sbgls}
\end{figure}

In Fig.\,\ref{fig:sbgls} we compare the sBGLS periodogram of the RV data where the dominating signals of planets b and c were removed with a simulation of a coherent signal of 172\,d period. A coherent 172\,d signal with properties taken from Table\,\ref{tab:modresult_c} was added to the residuals of the CARMENES RV data, where the five signals had been removed. 

\section{Very long-period variations}

\begin{figure}[h]
    \centering
    \includegraphics[width=0.49\textwidth]{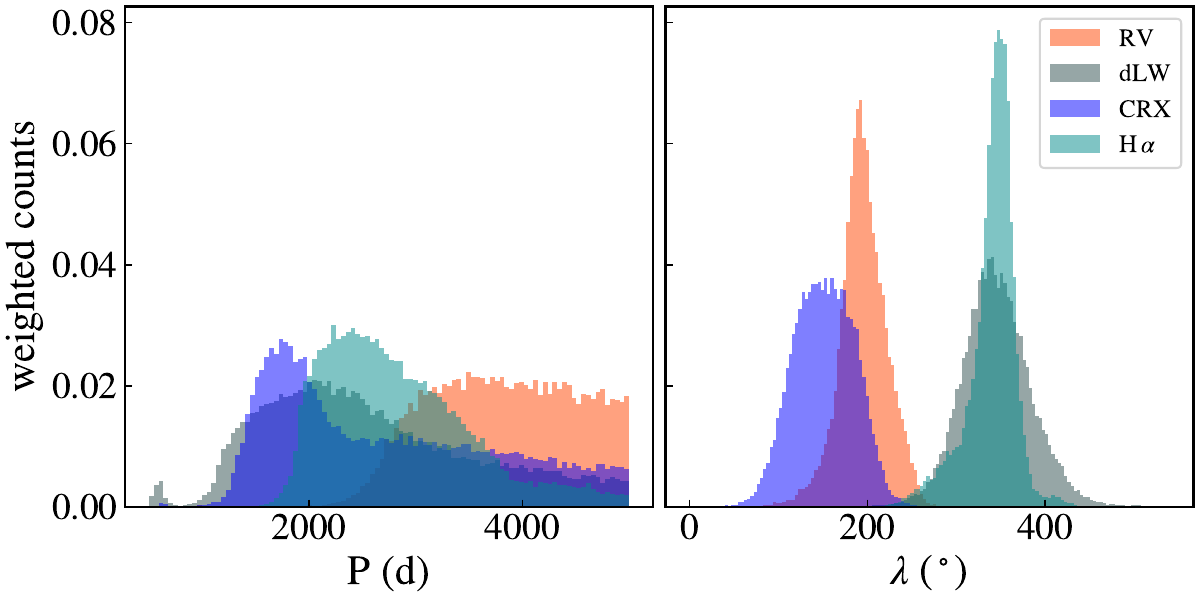}
    \caption{Posterior distributions for the orbital periods (left) and mean longitudes (right) for a fit of a Keplerian model of the CARMENES spectroscopic indices.}
    \label{fig:hist_cycle}
\end{figure}

The RV data and the spectroscopic activity indices show very long-period variations on the time scale of the total length of the dataset. We used identical priors ($K$: $\mathcal{U} [0,3]$\,m\,s$^{-1}$, $P$: $\mathcal{U} [500,5000]$\,d) to fit long-period signals and compare with the results illustrated by Fig.\,\ref{fig:hist_cycle}. Given the time base of our datasets of 2596\,d, the posterior distributions of the periods are consistent with each other. The phase shift between activity induced radial velocities and activity indicators is known for example from solar observations \citep{CollierCameron2019MNRAS.487.1082C}.

\section{Additional tables}

In Table\,\ref{tab:modresult_d} we list the priors and posteriors of the nine RV datasets, the jitter and offset parameters for each instrument as well as a linear and quadratic term. The two signals in the RV data, which are likely not due to orbiting planets, are listed in Table\,\ref{tab:modresult_c}. 

\begin{table}
    \centering
    \caption{Fit and derived dataset parameters from model E. The values for the dSHO kernel are from model I.}
    \begin{tabular}{@{}lr@{}ll@{}}
        \hline
        \hline
        \noalign{\smallskip}
         Parameter  & \multicolumn{2}{c}{Posterior}& Prior distribution\\
         \hline
         Jitter$_{\rm CARM VIS}$ [m\,s$^{-1}$] &$ 0.95$  & $^{+0.09}_{-0.05}$ & $\log \mathcal{U}$ [$10^{-8}$, $10^4$]\\
         Jitter$_{\rm ESPRESSO}$ [m\,s$^{-1}$] &$ 1.17$  & $^{+0.53}_{-0.22}$ & $\log \mathcal{U}$ [$10^{-8}$, $10^4$]\\
         Jitter$_{\rm MX R1}$    [m\,s$^{-1}$] &$ 1.05$  & $^{+0.28}_{-0.12}$ & $\log \mathcal{U}$ [$10^{-8}$, $10^4$]\\
         Jitter$_{\rm MX R2}$    [m\,s$^{-1}$] &$ 1.16$  & $^{+0.43}_{-0.20}$ & $\log \mathcal{U}$ [$10^{-8}$, $10^4$]\\
         Jitter$_{\rm MX R3}$    [m\,s$^{-1}$] &$ 1.39$  & $^{+0.55}_{-0.32}$ & $\log \mathcal{U}$ [$10^{-8}$, $10^4$]\\
         Jitter$_{\rm MX B1}$    [m\,s$^{-1}$] &$ 1.36$  & $^{+0.67}_{-0.34}$ & $\log \mathcal{U}$ [$10^{-8}$, $10^4$]\\
         Jitter$_{\rm MX B2}$    [m\,s$^{-1}$] &$ 1.49$  & $^{+0.86}_{-0.44}$ & $\log \mathcal{U}$ [$10^{-8}$, $10^4$]\\
         Jitter$_{\rm MX B3}$    [m\,s$^{-1}$] &$ 1.14$  & $^{+0.43}_{-0.19}$ & $\log \mathcal{U}$ [$10^{-8}$, $10^4$]\\
         Jitter$_{\rm HPF}$      [m\,s$^{-1}$] &$ 1.87$  & $^{+0.37}_{-0.33}$ & $\log \mathcal{U}$ [$10^{-8}$, $10^4$]\\
         Offset$_{\rm CARM VIS}$ [m\,s$^{-1}$] &$-1.24$  & $^{+0.21}_{-0.21}$ & $\mathcal{U}$ $[-9, 9]$\\
         Offset$_{\rm ESPRESSO}$ [m\,s$^{-1}$] &$ 1.33$  & $^{+0.81}_{-0.77}$ & $\mathcal{U}$ $[-9, 9]$\\
         Offset$_{\rm MX R1}$    [m\,s$^{-1}$] &$-1.04$  & $^{+0.71}_{-0.75}$ & $\mathcal{U}$ $[-9, 9]$\\
         Offset$_{\rm MX R2}$    [m\,s$^{-1}$] &$-0.91$  & $^{+0.98}_{-0.91}$ & $\mathcal{U}$ $[-9, 9]$\\
         Offset$_{\rm MX R3}$    [m\,s$^{-1}$] &$-0.66$  & $^{+1.07}_{-1.17}$ & $\mathcal{U}$ $[-9, 9]$\\
         Offset$_{\rm MX B1}$    [m\,s$^{-1}$] &$-1.31$  & $^{+0.95}_{-0.99}$ & $\mathcal{U}$ $[-9, 9]$\\
         Offset$_{\rm MX B2}$    [m\,s$^{-1}$] &$-0.68$  & $^{+1.20}_{-1.15}$ & $\mathcal{U}$ $[-9, 9]$\\
         Offset$_{\rm MX B3}$    [m\,s$^{-1}$] &$-0.86$  & $^{+1.09}_{-1.20}$ & $\mathcal{U}$ $[-9, 9]$\\
         Offset$_{\rm HPF}$      [m\,s$^{-1}$] &$ 0.48$  & $^{+0.39}_{-0.39}$ & $\mathcal{U}$ $[-9, 9]$\\
         Linear trend           [m\,s$^{-1}$\,yr$^{-1}$] &$0.22$& $^{+0.12}_{-0.12}$&$\mathcal{U}$ $[-0.73,0.73]$\\
         Quad. term        [m\,s$^{-1}$\,yr$^{-2}$] &$ 0.22$ & $^{+0.05}_{-0.05}$&$\mathcal{U}$ $[-0.4,0.4]$\\
        \noalign{\medskip}
        $\sqrt\sigma_{\rm dSHO}$ [m$^1$\,s$^{-1}$] &1.3  & $^{+0.5}_{-0.3}$ & $\log \mathcal{U}$ $[0.001, 10]$\\
        P$_{\rm dSHO}$      [d]               &175.1& $^{+2.7}_{-2.6}$ & $\log \mathcal{U}$ $[150, 200]$\\
        Q$_{\rm dSHO}$                        &23 & $^{+75}_{-18}$ & $\mathcal{U}$ $[0.01, 100]$\\
        dQ$_{\rm dSHO}$                       &56 & $^{+30}_{-35}$ &$\mathcal{U}$ $[0.01, 100]$\\
        f$_{\rm dSHO}$                        &1.4& $^{+2.7}_{-1.0}$ & $\mathcal{U}$ $[0, 10]$\\
         \hline
    \end{tabular}
    \label{tab:modresult_d}
\end{table}

\begin{table}[t]
    \centering
    \caption{Fit and derived parameters for the signals at 96\,d and 172\,d represented by Keplerians. }
    \begin{tabular}{@{}lr@{}ll@{}}
        \hline
        \hline
        \noalign{\smallskip}
         Parameter  & \multicolumn{2}{c}{Posterior}& Prior distribution\\
         \hline
         \multicolumn{4}{c}{$P$ = 172\,d}\\
         $P$ [d]                &$ 172.5$ & $^{+1.5}_{-1.3}$   & $\mathcal{U}\, [160,190]$\\
         $K$ [m\,s$^{-1}$]              &$ 1.30 $ &$^{+0.20}_{-0.20}$  & $\mathcal{U}\, [0,3.5]$\\
         $h$                    &$ 0.07 $ & $^{+0.24}_{-0.25}$ & $\mathcal{N}\, [0,0.45, -1, 1]$\\
         $k$                    &$-0.06 $ & $^{+0.23}_{-0.24}$ & $\mathcal{N}\, [0,0.45, -1, 1]$\\
         $\lambda$ [deg] &$ 300.1$ & $^{+9.7}_{-10.2}$  & $\mathcal{U}\, [0,360]$\\
        \noalign{\smallskip}
         \hline
         \multicolumn{4}{c}{$P$ = 96\,d}\\
         $P$ [d]                & 96.1  & $^{+0.9}_{-0.6}$  & $\mathcal{U}\, [88,110]$\\
         $K$ [m/s]              & 1.09  &$^{+0.36}_{-0.27}$ & $\mathcal{U}\, [0,2]$\\
         $h$                    & 0.29  & $^{+0.25}_{-0.34}$& $\mathcal{N}\, [0,0.45, -1, 1]$\\
         $k$                    & 0.43  & $^{+0.22}_{-0.36}$& $\mathcal{N}\, [0,0.45, -1, 1]$\\
         $\lambda$ [deg] & 333 & $^{+13}_{-15}$      & $\mathcal{U}\, [0,360]$\\
        \noalign{\smallskip}
         \hline
    \end{tabular}
    \label{tab:modresult_c}
\end{table}

\section{Corner plot}

\begin{figure*}
    \centering
    \includegraphics[width=0.99\textwidth]{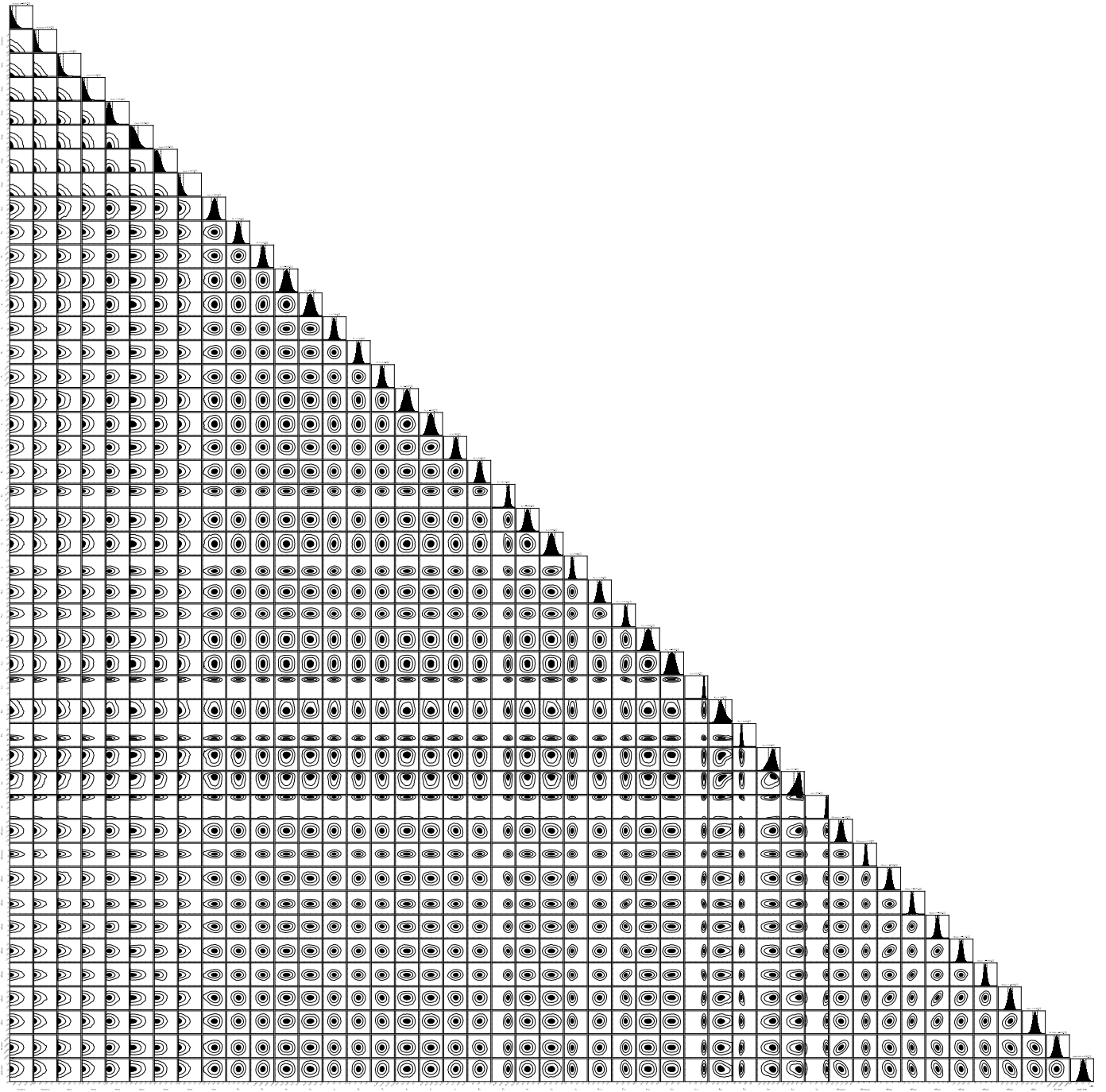}
    \caption{Corner plot of all fit parameters. From left to right: Jitter terms for all instruments, Keplerian parameters for 5 signals, RV offsets for all instruments as well as the linear and quadratic trend terms.}
    \label{fig:corner}
\end{figure*}

The corner plot of the best model (model E from Table\,\ref{tab:modselect}) is shown in Fig.\,\ref{fig:corner}. It depicts the posteriors and the covariances of all parameters used. The results are also listed in Tables\,\ref{tab:modresult_p}, \ref{tab:modresult_d}, and \ref{tab:modresult_c}.

\begin{figure}
    \centering
    \includegraphics[width=0.49\textwidth]{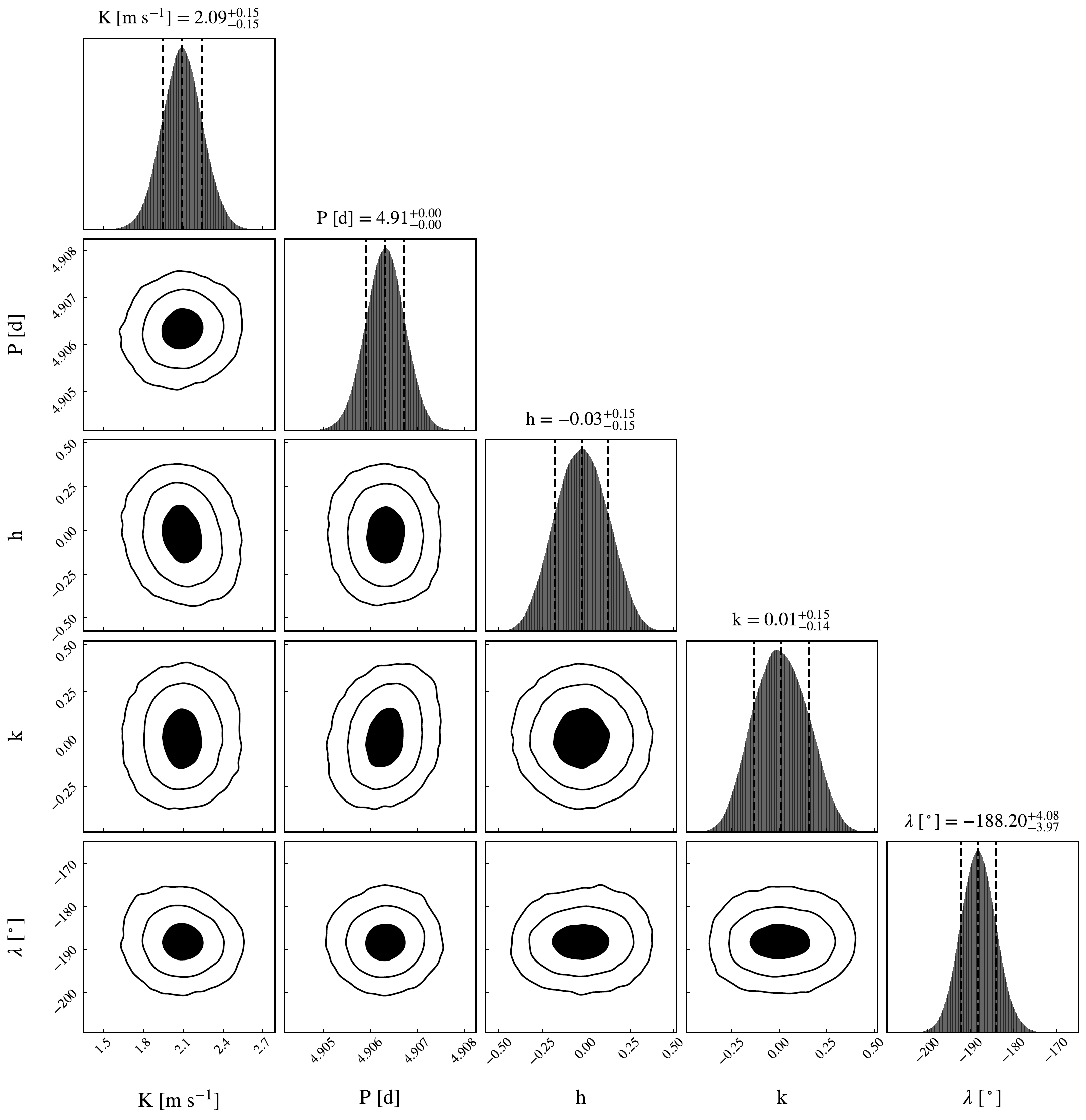}
    \caption{Corner plot of parameters for planet b.}
    \label{fig:corner_b}
\end{figure}

\begin{figure}
    \centering
    \includegraphics[width=0.49\textwidth]{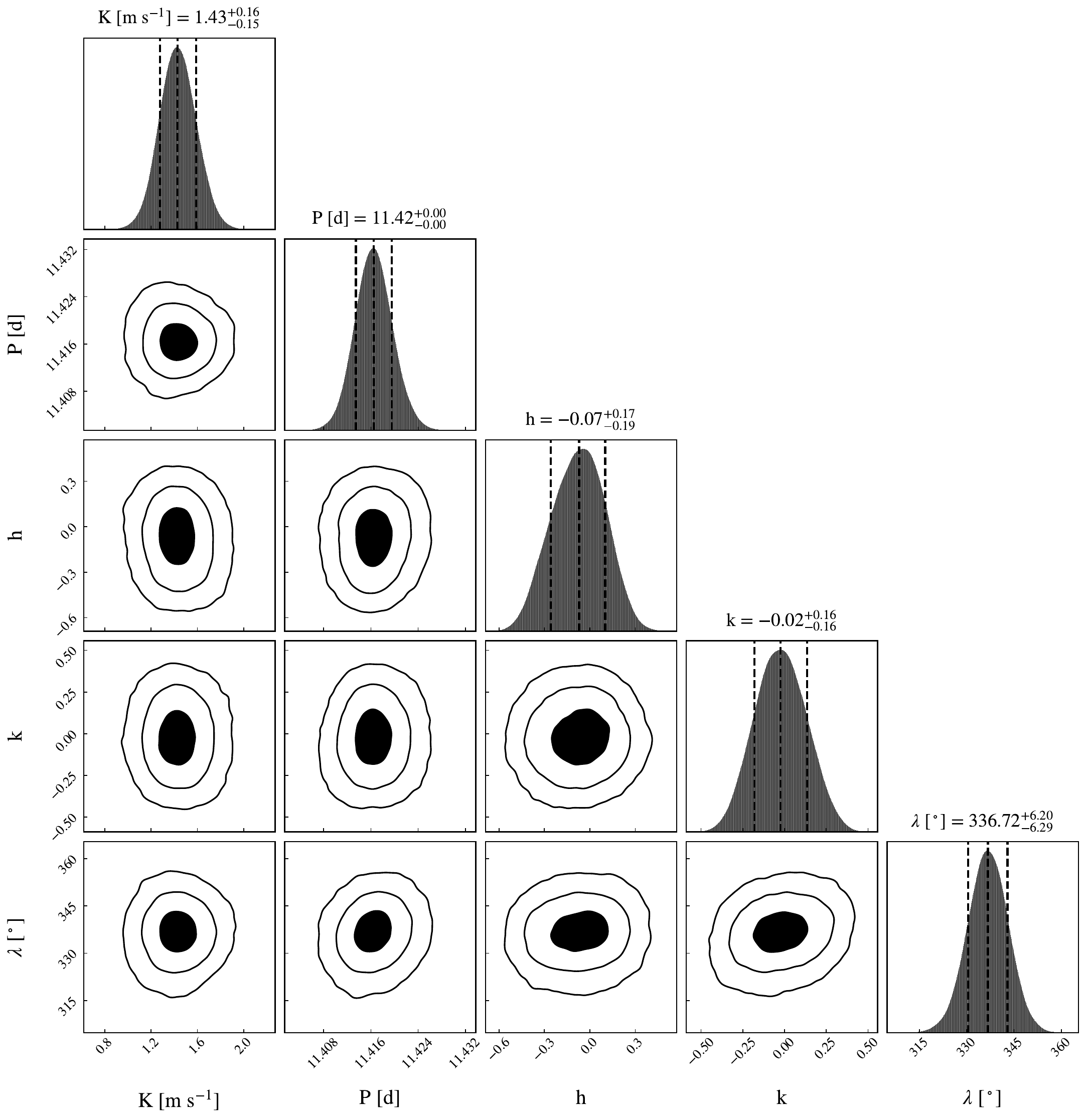}
    \caption{Corner plot of parameters for planet c.}
    \label{fig:corner_c}
\end{figure}

\begin{figure}
    \centering
    \includegraphics[width=0.49\textwidth]{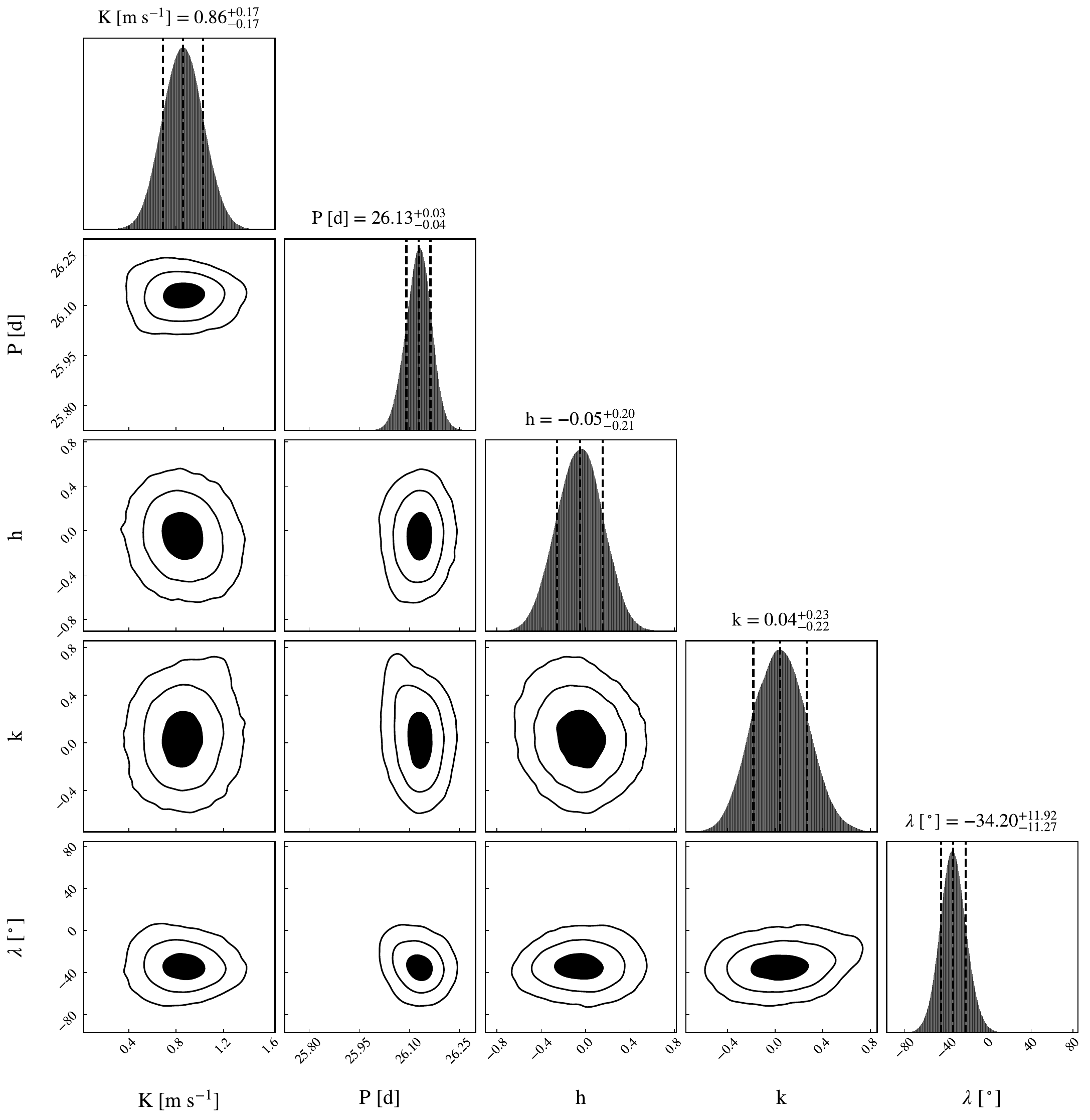}
    \caption{Corner plot of parameters for planet d.}
    \label{fig:corner_d}
\end{figure}

\section{TESS transit detection}

\begin{figure}
    \centering
    \includegraphics[width = 0.49\textwidth]{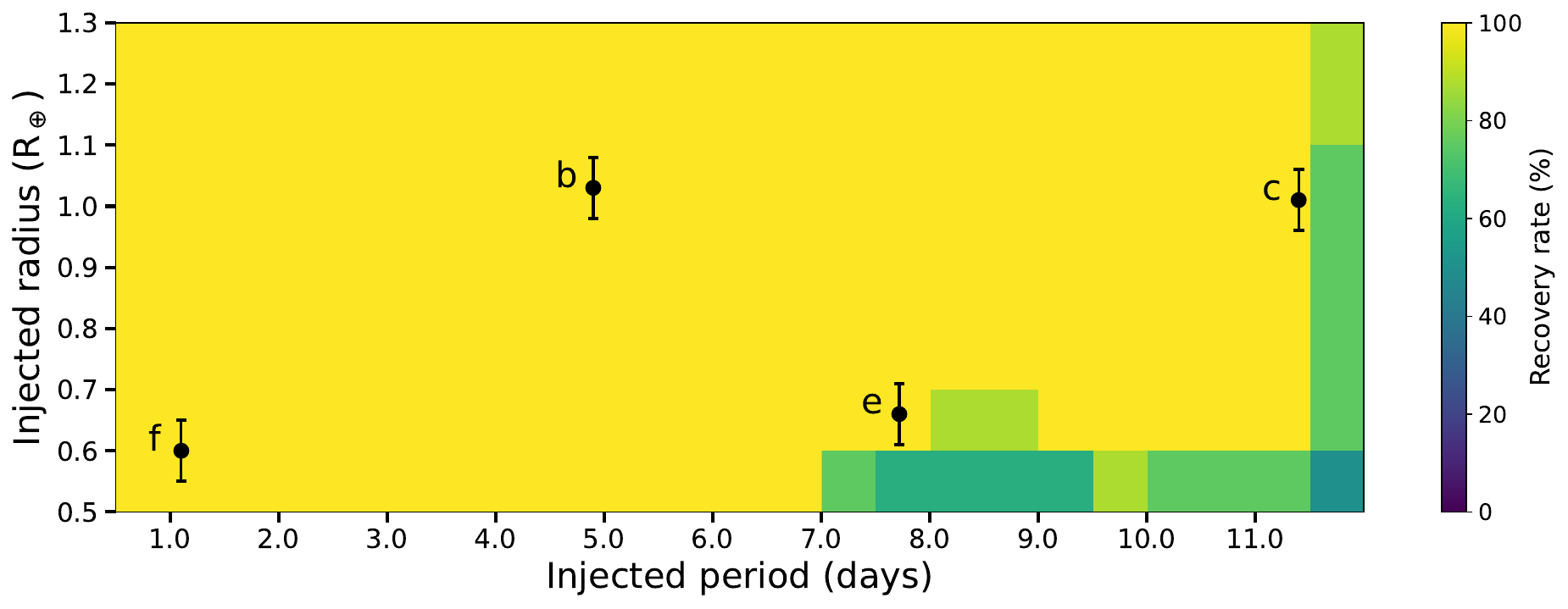}
    \caption{Planet injection and recovery test for the planets and potential planet candidates with periods below 12\,d using {\tt MATRIX} and {\em TESS} sector 43.}
    \label{fig:recovery}
\end{figure}

\begin{figure}
    \centering
    \includegraphics[width = 0.49\textwidth]{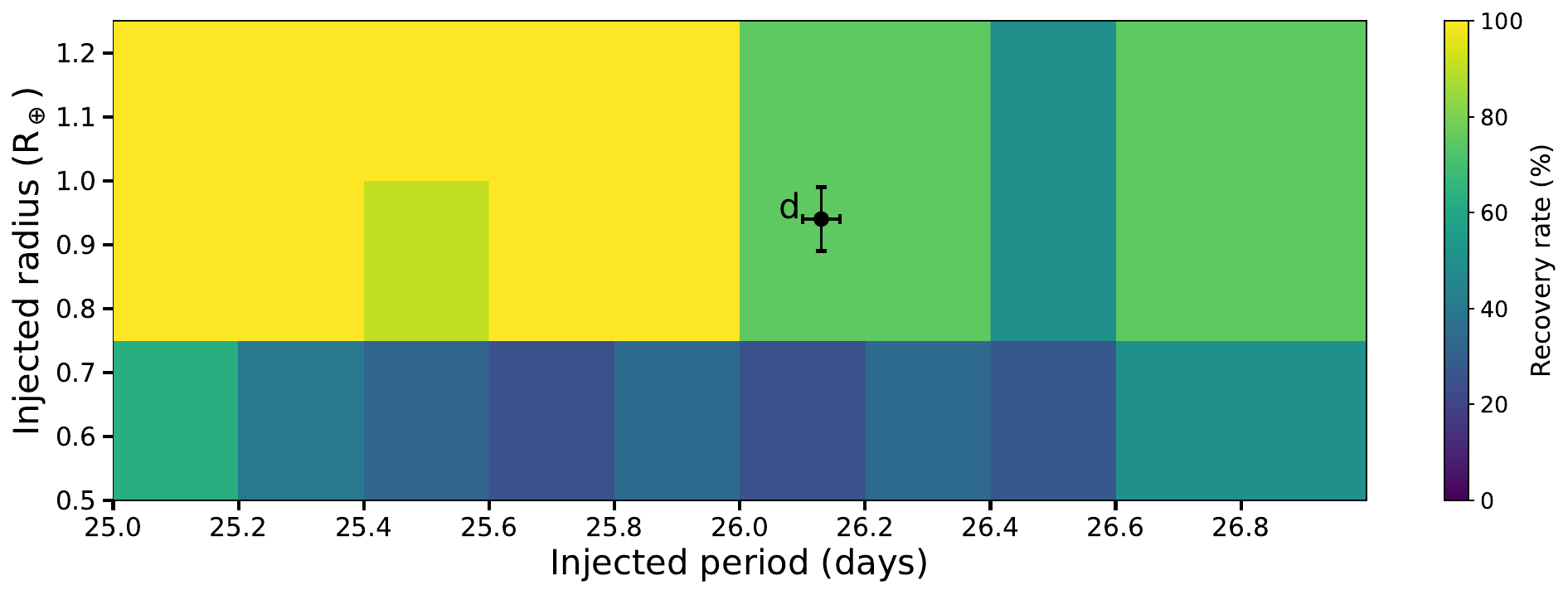}
    \caption{Planet injection and recovery test for the planet d using {\tt MATRIX} and {\em TESS} sectors 70 and 71.}
    \label{fig:recovery_26}
\end{figure}

Here we present the result of the planet injection test using {\tt MATRIX} \footnote{{The {\tt MATRIX (\textbf{M}ulti-ph\textbf{A}se \textbf{T}ransits \textbf{R}ecovery from \textbf{I}njected e\textbf{X}oplanets}) code is open access on GitHub: \url{https://github.com/PlanetHunters/tkmatrix}}} for {\em TESS} sectors 43, 70, and 71. Planets with orbital periods between 1\,d and 12\,d and radii between 0.5\,R$_\oplus$ and 1.3\,R$_\oplus$ were injected in {\em TESS} sector 43 with four different phase values. The recovery fraction is shown in Fig.\,\ref{fig:recovery}. The potential planet labeled e and f corresponds to the signal at 7.2\,d and 1.1\,d (models F and G) discussed in Section 4.4.1. Using the two consecutive sectors 70 and 71, we tested the detectability of planet\,d injecting planets with periods between 25\,d and 27\,d and radii between 0.5\,R$_\oplus$ and 1.3\,R$_\oplus$ (Fig.\,\ref{fig:recovery_26}). The planet radii shown in the two figures assume Earth density, the assumed masses are the minimum masses and the corresponding uncertainty. 


\end{appendix}
\end{document}